\documentclass[12pt,aps,pre,preprint,tightenlines,floatfix,showpacs,superscriptaddress,raggedbottom]{revtex4-1}
\usepackage{amssymb,amsmath,bm,hyperref,graphicx}
\begin{document}

\title{Stepwise introduction of model complexity in a generalized master equation
                      approach to time-dependent transport}

\author{Vidar Gudmundsson}
\email{vidar@hi.is}
\affiliation{Science Institute, University of Iceland, Dunhaga 3, IS-107 Reykjavik, Iceland}

\author{Olafur Jonasson}
\affiliation{Science Institute, University of Iceland, Dunhaga 3, IS-107 Reykjavik, Iceland}

\author{Thorsten Arnold}
\affiliation{Science Institute, University of Iceland, Dunhaga 3, IS-107 Reykjavik, Iceland}

\author{Chi-Shung Tang}
\email{cstang@nuu.edu.tw}
\affiliation{Department of Mechanical Engineering, National United University, 1, Lienda, Miaoli 36003, Taiwan}

\author{Hsi-Sheng Goan}
\email{goan@phys.ntu.edu.tw}
\affiliation{Department of Physics and Center for Theoretical Sciences, 
                   National Taiwan University, and\\
                   Center for Quantum Science and Engineering, 
                   National Taiwan University, Taipei 10617, Taiwan}

\author{Andrei Manolescu}
\email{manoles@ru.is}
\affiliation{Reykjavik University, School of Science and Engineering, Menntavegur 1, IS-101 Reykjavik, Iceland}

\begin{abstract}
      We demonstrate that with a stepwise introduction of complexity 
      to a model of an electron system embedded in a photonic cavity and a 
      carefully controlled stepwise truncation of the ensuing many-body space
      it is possible to describe the time-dependent transport of electrons 
      through the system with a non-Markovian generalized quantum master
      equation. We show how this approach retains effects of an external magnetic field
      and the geometry of an anisotropic electronic system. The Coulomb interaction between 
      the electrons and the full electromagnetic coupling between the electrons and the photons are 
      treated in a non-perturbative way using ``exact numerical diagonalization''.   
\end{abstract}

\maketitle

\section{Introduction}
Advances in techniques constructing and experimenting with 
quantum electrodynamic circuits have resulted in systems with very strong electron-photon
coupling \cite{Niemczyk10:772,Frey11:01,Delbecq11:01}. Traditionally, some version of the
Jaynes-Cummings model \cite{Jaynes63:89} is used to describe the energy spectrum of the 
closed system or its time evolution \cite{Amitabh93:2276}. Recently, we have shown that
for a strong electron-photon coupling in a semiconductor nanostructure the Jaynes-Cumming 
model may not be adequate and one may have to consider a model with more than two electron 
levels and the diamagnetic term in the coupling \cite{Jonasson2011:01}. 
In continuation we have used our experience with describing the dynamics of open systems
in terms of the generalized master equation (GME) \cite{Moldoveanu10:155442,Gudmundsson10:205319}
to start the exploration of time-dependent transport properties of circuit quantum electrodynamic
(circuit-QED) systems \cite{Gudmundsson12:1109.4728}.

Here, we will describe our approach with a special emphasis on what we call: 
``A stepwise introduction of complexity to a model and a carefully controlled stepwise truncation 
of the ensuing many-body space''. We will discuss technical issues that are common to
models of different phenomena and fields, but we will use the model of Coulomb interacting electrons
in a photonic cavity as an example to display our approach and findings. 
\begin{figure}[htbq]
      \begin{center}
            \includegraphics[width=0.80\textwidth,angle=0,bb=1 71 675 355,clip]{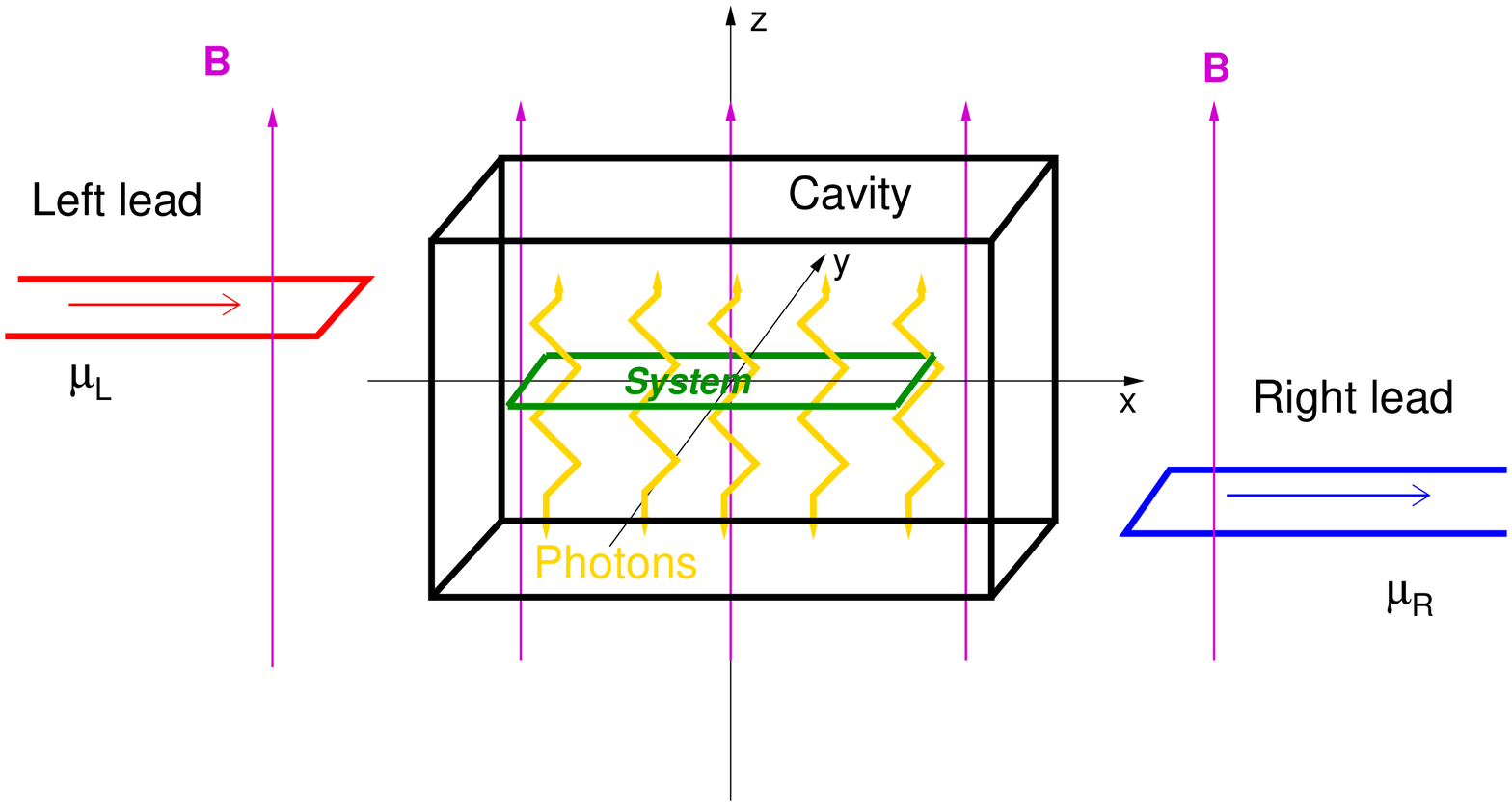}
      \end{center}
      \caption{A schematic view of the total system consisting of a finite 
               quantum wire (green, the central electronic system) in a photon
               cavity (bold black), coupled to external left (red) and right (blue) leads 
               at different chemical potentials, $\mu_\mathrm{L}$ and
               $\mu_\mathrm{R}$. An external homogeneous magnetic field 
               $\mathbf{B}$ (magenta) is perpendicular to the system and
               the leads. The lengths are not shown to scale. The length of the central
               system is 300 nm, but the characteristic lengths of the cavity are on the 
               millimeter scale. The GaAs leads and central system contain a 
               quasi-one-dimensional (Q1D) electron gas in 1 to 4 subbands 
               of width 60 to 120 nm for the parameters to be introduced below. The cavity
               photon modes (yellow) are standing waves in the $z$-direction with an electric
               component in the $x$- or $y$-direction. The applied bias between the left and
               right leads is indicated by placing them at different levels with respect to 
               the central system.}
      \label{L-S}
\end{figure}

\section{The model of the closed system}
The closed electron-photon system is a finite quasi-one-dimensional (Q1D) quantum wire 
placed in the center of a rectangular photon cavity (see Figure \ref{L-S}). 
It is described by the Hamiltonian \cite{Gudmundsson12:1109.4728} 
\begin{align}
        H&_0=\sum_i E_id_i^{\dagger}d_i + \hbar\omega a^{\dagger}a +
        \frac{1}{2}\sum_{ijrs}\langle ij|V_{\mathrm{Coul}}|rs\rangle 
        d_i^{\dagger}d_j^{\dagger}d_sd_r \nonumber\\ \label{H-e-EM}
        &+{\cal E}_\mathrm{c}\sum_{ij}d_i^{\dagger}d_j\; g_{ij}\left\{a + a^\dagger\right\}\\
        &+{\cal E}_\mathrm{c}\left(\frac{{\cal E}_\mathrm{c}}{\hbar\Omega_w}\right)
        \sum_{i}d_i^{\dagger}d_i\left\{\left( a^\dagger a+\frac{1}{2}\right) +
        \frac{1}{2}\left( aa+a^\dagger a^\dagger\right)\right\}{\nonumber},
\end{align}
where $E_i$ is the single-electron spectrum for the finite quantum wire with hard
walls at $x=\pm L_x/2$ and parabolic confinement in the $y$-direction with characteristic 
energy $\hbar\Omega_0$. A static classical external magnetic field  
$\mathbf{B}=\mathbf{\nabla}\times\mathbf{A}_{\mathrm{ext}}=B\hat{\mathbf{z}}$
renormalizes the frequency of the $y$-confinement $\Omega_w^2=\omega_c^2+\Omega_0^2$
and the natural length scale $a_w=\sqrt{\hbar /(m^*\Omega_w)}$ with the cyclotron 
frequency $\omega_c=eB/(m^*c)$. $d_i$ is an annihilation operator of 
the non-interacting single-electron state (SES) $|i\rangle$ with energy $E_i$, and $a$ is the
annihilation operator for the single-photon mode with energy $\hbar\omega$.
The kernel for the Coulomb interaction of the electrons is
\begin{equation}
      V_{\mathrm{Coul}}(\mathbf{r}-\mathbf{r}')=
      \frac{e^2}{\kappa\sqrt{(x-x')^2+(y-y')^2+\eta^2}},
\label{Coul}
\end{equation}
with the small regularization factor selected such that 
$\eta /a_w\approx 7.1\times 10^{-3}$ when $a_w\approx 33.5$ nm at
$B=0.1$ T for GaAs parameters, i.\ e.\ the effective mass $m^*=0.067m_e$, 
and the dielectric constant $\kappa =12.4$.
The second line in the Hamiltonian (\ref{H-e-EM}) is the paramagnetic interaction
between electrons and photons $(-\int d{\mathbf{r}}\;\mathbf{j}\cdot\mathbf{A})/c$, 
and the last line stems from the diamagnetic term in the interaction
$(-e\int d{\mathbf{r}}\;\rho A^2)/(2m^*c^2)$. In terms of the field operators 
the charge current density and charge density are, respectively,
\begin{equation}
      \mathbf{j} = -\frac{e}{2m^*}\left\{\psi^\dagger\left({\bm{\pi}}\psi\right)
                               +\left({\bm{\pi}}^*\psi^\dagger\right)\psi\right\},
\end{equation}
and
\begin{equation}
      \rho = -e\psi^\dagger\psi, 
\end{equation}
where
\begin{equation}
      {\bm{\pi}}=\left(\mathbf{p}+\frac{e}{c}\mathbf{A}_{\mathrm{ext}}\right).
\end{equation}

The photon cavity is assumed to be a rectangular box $(x,y,z)\in\{[-a_\mathrm{c}/2,a_\mathrm{c}/2]
\times [-a_\mathrm{c}/2,a_\mathrm{c}/2]\times [-d_\mathrm{c}/2,d_\mathrm{c}/2]\}$ 
with the finite quantum wire centered in the $z=0$ plane. In the Coulomb gauge the
polarization of the electric field can be chosen parallel to the transport
in the $x$-direction by selecting the TE$_{011}$ mode, or perpendicular to it by selecting the 
TE$_{101}$ mode 
\begin{equation}
      \mathbf{A}(\mathbf{r})=\left({\hat{\mathbf{e}}_x\atop \hat{\mathbf{e}}_y}\right)
      {\cal A}\left\{a+a^{\dagger}\right\}
      \left({\cos{\left(\frac{\pi y}{a_\mathrm{c}}\right)}\atop\cos{\left(\frac{\pi x}{a_\mathrm{c}}\right)}} \right)
      \cos{\left(\frac{\pi z}{d_\mathrm{c}}\right)}, \quad\quad
      {\mbox{TE$_{011}$}\atop\mbox{TE$_{101}$}}. 
\label{Cav-A}
\end{equation}
${\cal A}$ is the amplitude of the cavity vector field defining a characteristic
energy scale ${\cal E}_\mathrm{c}=e{\cal A}\Omega_wa_w/c = g^\mathrm{EM}$ for the
electron-photon interaction and leaving an effective dimensionless coupling tensor
\begin{equation}
\label{g_ab}
      g_{ij} = \frac{a_w}{2\hbar}\int d{\mathbf{r}}\; 
      [\psi_i^*(\mathbf{r})\left\{\left(\hat{\mathbf{e}}\cdot\bm{\pi}\right)
      \psi_j(\mathbf{r})\right\}
      +\left\{\left(\hat{\mathbf{e}}\cdot\bm{\pi}\right)
      \psi_i(\mathbf{r})\right\}^*\psi_j(\mathbf{r}) ],
\end{equation}
defining the coupling of individual single-electron states $|i\rangle$ and 
$|j\rangle$ to the photonic mode. In the calculations of the energy spectrum of 
the Hamiltonian (\ref{H-e-EM}) we will retain all resonant and antiresonant terms 
in the photon creation and annihilation operators so we will not use
the rotating wave approximation, but in the calculations of the electron-photon
coupling tensor (\ref{g_ab}) we assume $a_w,L_x << a_\mathrm{c}$ and approximate
$\cos(\pi\{x,y\}/a_\mathrm{c})~\sim 1$ in Eq.\ (\ref{Cav-A}) for the cavity vector 
field $\mathbf{A}$. 

The energy spectrum and the states of the Hamiltonian for the closed electron-photon 
system have to be sought for an unspecified number of electrons as we want to open the 
system up for electrons from the leads later. We will be investigating systems with
few electrons present in the finite quantum wire, but it is not a trivial task to
construct an adequate many-body (MB) basis for the diagonalization of $H_0$ since in addition
to geometrical and bias (set by the leads) considerations we have strong requirements
set both by the Coulomb and the photon interaction. Our solution to this dilemma and
a mean to keep a tight lid on the exponential growth of the size of the required 
many-body Fock space is to do the diagonalization in two steps.

First, we select the lowest $N_{\mathrm{SES}}$ single-electron states (SESs) of the 
finite quantum wire. These have been found by diagonalizing the Hamiltonian operator
for a single electron in the Q1D confinement and in a perpendicular constant magnetic field in a large 
basis of oscillator-like wave functions. 
Originally, we constructed a many-electron Fock space with $N_\mathrm{MES}
=2^{N_{\mathrm{SES}}}$ states $|\mu\rangle$ \cite{Moldoveanu09:073019,Gudmundsson09:113007}. 
(We use Latin indices for the single-electron states and Greek ones for the many-electron states). 
This ``simple binary'' construction for the Fock-space does not deliver the optimal ratio
of single-, two-, and higher number-of-electrons states for an interacting system when
their energy is compared. Usually one ends up with too few SESs compared
to the MESs. Here we will select 18 SESs and construct all possible combinations of 2-4 electron states.
This can be refined further. These MESs, which are in fact Slater determinants, and which we denote
as $|\mu\rangle$ (with an angular right bracket) are then used to diagonalize the 
part of the Hamiltonian (\ref{H-e-EM}) for the Coulomb interacting electrons only,
supplying their spectrum ${\tilde E}_\mu$ and states $|\mu )$ (denoted now with a rounded 
right bracket). The eigenvectors from
the diagonalization procedure define the unitary transform between the two sets of
MESs, $|\mu ) = \sum_{\alpha}{\cal V}_{\mu\alpha}|\alpha\rangle$. As the action of the 
creation and annihilation operators is only known in the non-interacting electron
basis $\{|\mu\rangle\}$ we need this transform to write $H_0$ in the new Coulomb
interacting basis $\{|\mu )\}$
\begin{align}
      H_0&=\sum_\mu |\mu )\tilde{E}_\mu(\mu| + \hbar\omega a^{\dagger}a  
      + {\cal E}_\mathrm{c}\sum_{\mu\nu ij}|\mu )
      \langle\mu |{\cal V}^+ d_i^{\dagger}d_j{\cal V}|\nu\rangle (\nu |\;
      g_{ij}\left\{a+a^\dagger\right\} \nonumber\\
      &+{\cal E}_\mathrm{c}\left(\frac{{\cal E}_\mathrm{c}}{\hbar\Omega_w}\right)
      \sum_{\mu\nu i}
      |\mu )\langle\mu |{\cal V}^+ d_i^{\dagger}d_i{\cal V}|\nu\rangle (\nu |   
      \left\{\left( a^\dagger a+\frac{1}{2}\right) +
      \frac{1}{2}\left( aa+a^\dagger a^\dagger\right)\right\}.
\label{H-e-VEMV}
\end{align}
In order to finally obtain the energy spectrum of the electron-photon Hamiltonian (\ref{H-e-VEMV})
we need to construct a MB-space 
$|\mu )\otimes|N_{\mathrm{ph}}\rangle\longrightarrow |\mu\rangle_{\mathrm{e-EM}}$ out of the 
Coulomb interacting MESs $|\mu )$ and the eigenstates of the photon number operator 
$|N_{\mathrm{ph}}\rangle$. To properly take account of the effects of the Coulomb interaction
we selected a large basis $\{|\mu\rangle\}$ and many photon states since the strong coupling to 
the cavity photons requires many states. 
For the system parameters to be introduced later we find that
basis build up of the 64 lowest in energy Coulomb interacting MESs and 27 photon states is 
adequate for the transport bias windows and the electron-photon coupling to be selected later. 
The catch is that the unitary transform has to be performed with the full untruncated basis
since ${\cal V}$ can not be truncated. 
In other words each of the 64 interacting MESs $|\mu )$ remains a superposition of all 
of the $2^{N_\mathrm{SES}}$ noninteracting MESs $|\mu\rangle$ with the same number of 
electrons $N_\mathrm{e}$, 
i.\ e.\ $|\mu )$ is a linear combination of a subset of $\binom{N_\mathrm{SES}}{N_\mathrm{e}}$ terms.
(A similar issue is met when the dimensionless coupling
tensor $g_{ij}$ (\ref{g_ab}) is transformed from the original single-electron basis to the 
single-electron states $|i\rangle$). 

The diagonalization of $H_0$ produces the new interacting electron-photon states
$|\breve{\mu}) = \sum_{\alpha}{\cal W}_{\mu\alpha}|\alpha\rangle_{\mathrm{e-EM}}$ with
a known integer electron content, but an indefinite number of photons, since the photon number 
operator does not commute with $H_0$. 
\begin{figure}[htbq]
      \begin{center}
            \includegraphics[width=0.325\textwidth,angle=0,bb=50 59 180 325,clip]{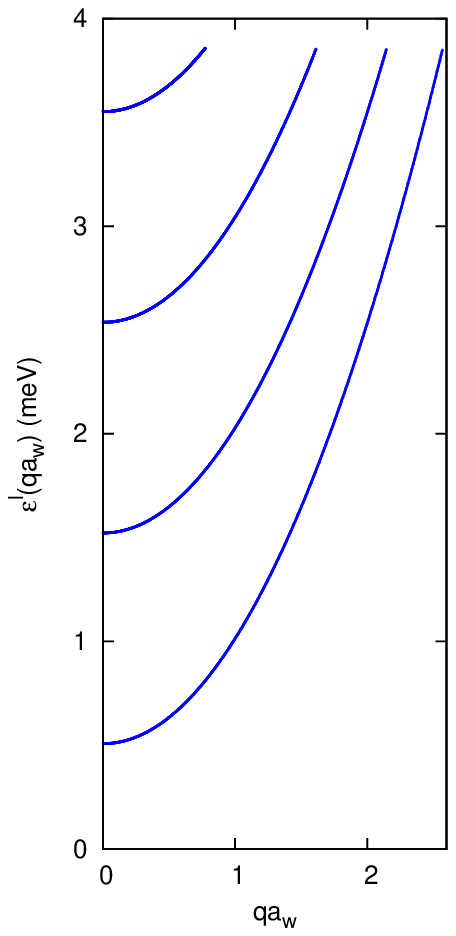}
            \includegraphics[width=0.32\textwidth,angle=0,bb=1 1 243 501,clip]{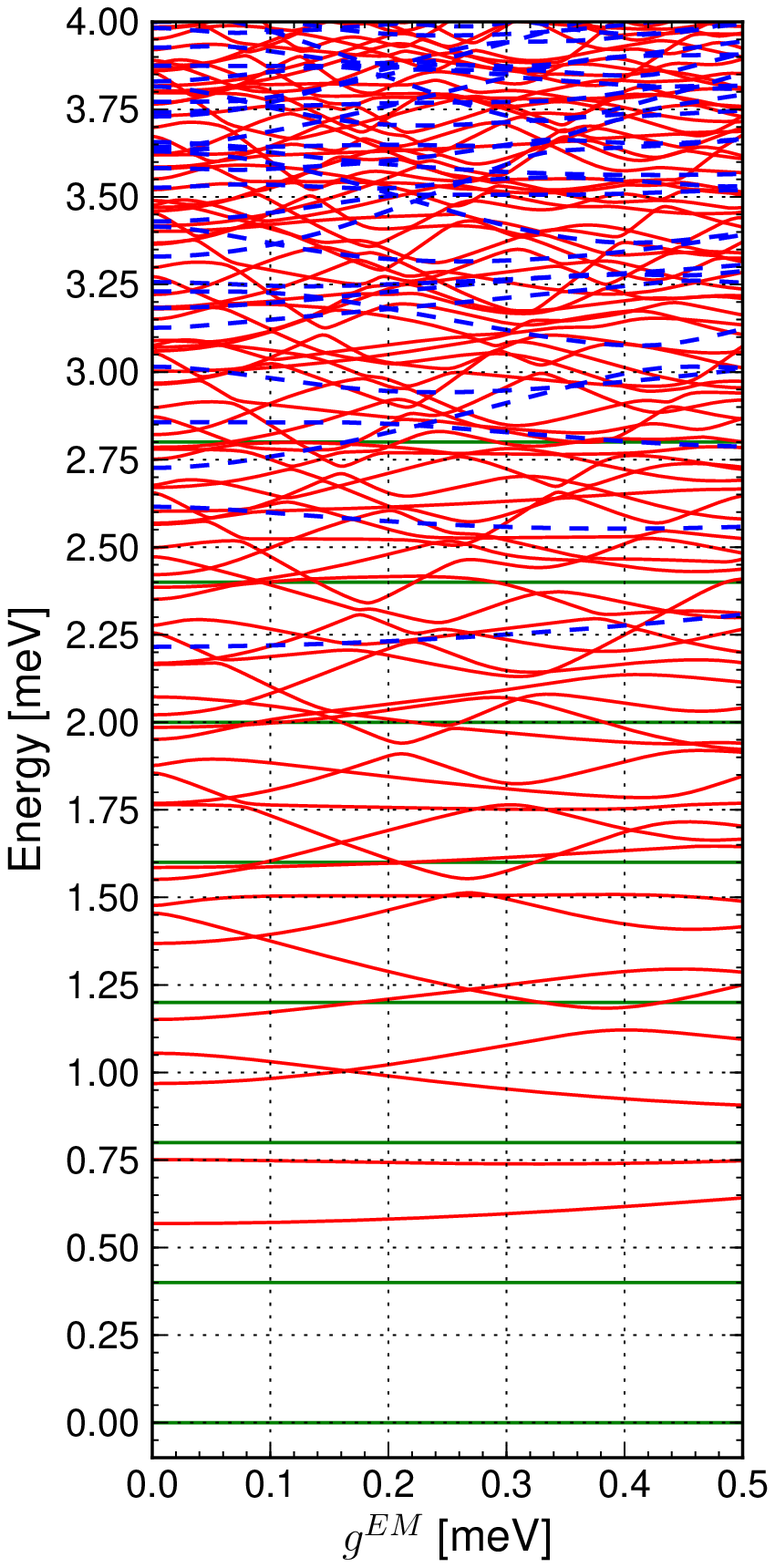}
            \includegraphics[width=0.32\textwidth,angle=0,bb=1 1 243 501,clip]{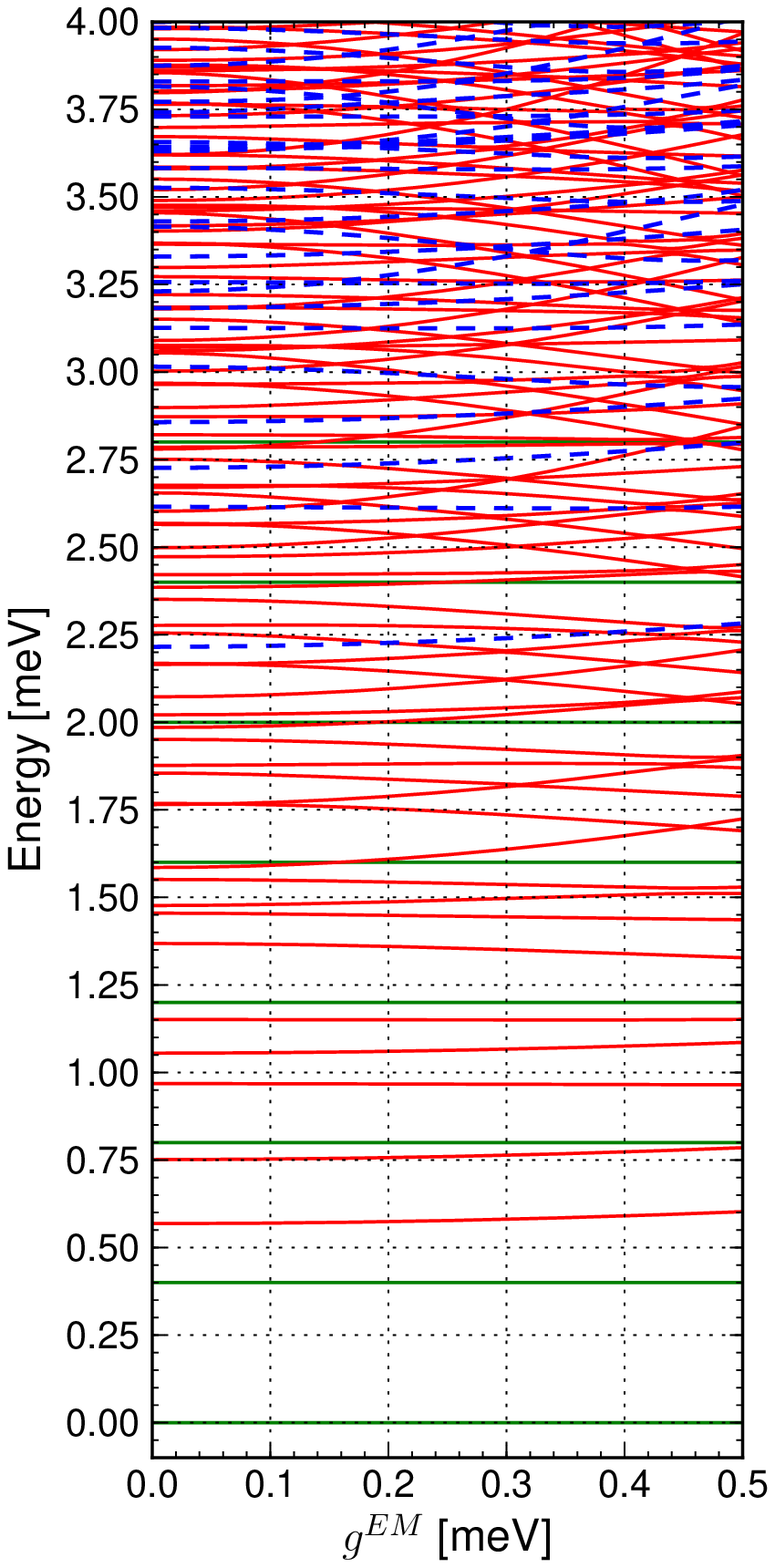}
      \end{center}
      \caption{The single-electron energy spectrum of the leads (left panel)
               versus the ``subband momentum'' $qa_w$, and
               the many-body energy spectra for Coulomb interacting electrons
               coupled to quantized cavity photon modes with the electric component polarized
               along the finite quantum wire (x-polarization, center panel), and perpendicular to the
               wire (y-polarization, right panel) versus the electron-photon coupling strength 
               $g^\mathrm{EM}={\cal E}_\mathrm{c}$. In the energy range shown there are 
               states with no electrons (horizontal solid green), one electron (solid red), 
               and two electrons (dashed blue). In order to reach convergence for high
               ${\cal E}_\mathrm{c}$ in this figure we use $N_\mathrm{SES}=200$ here.
               $B=0.1$ T, $\hbar\Omega_0=1.0$ meV, $\hbar\omega = 0.4$ meV, $\hbar\Omega_0^l=1.0$ meV,
               $L_x=300$ nm, $m^*=0.067m_e$, $\kappa =12.4$, the dielectric constant of GaAs.}
      \label{EB-e-EM-rof}
\end{figure}
The MB energy spectra for the $x$- and $y$-polarization of the photon mode are shown in
Fig.\ \ref{EB-e-EM-rof} in comparison with the subbandstructure of the 
single-electron spectrum of the leads. The difference in the energy spectra for the $x$- and the $y$-polarization
stems from the anisotropic electron system. The photon mode selected with energy $0.4$ meV is 
close to be in resonance with the motion in the $x$-direction, but quite far from the fundamental
energy in the $y$-direction, $1.0$ meV. The spectrum for the $x$-polarization thus displays
stronger dispersion and interaction of levels than the spectrum for the $y$-polarization.

The states of the closed system in Fig.\ \ref{EB-e-EM-rof} have a definite
number of electrons and an undetermined number of photons of one given polarization.
Later when the system is coupled to the leads one can expect the electrons entering the 
system to radiate photons with any polarization. A preferred polarization could though
be influenced by the aspect ratio of the cavity. In the treatment of the time-dependent
system to follow we assume the electrons only to radiate photons with one given polarization.
This is done to facilitate the numerical calculations and the analysis of a system with a
strong spatial anisotropy.

\section{Opening of the system, coupling to external leads}
At time $t=0$ the closed system of Coulomb interacting electrons coupled to 
cavity photons described by the Hamiltonian (\ref{H-e-VEMV}) 
is connected to two external semi-infinite quasi-one-dimensional
quantum wires in a perpendicular magnetic field \cite{Gudmundsson10:205319}. 
We use an approach introduced by Nakajima and Zwanzig to project the time 
evolution of the total system onto the
central system by partial tracing operations with respect to the operators
of the leads \cite{Nakajima58:948,Zwanzig60:1338}. The coupling Hamiltonian of the
central system (i.\ e.\ the short Q1D wire) to the leads is
of the form
\begin{equation}
      H_\mathrm{T}(t)=\sum_{i,l}\chi^l(t)\int dq\;
      \left\{T^l_{qi}c_{ql}^\dagger d_i + (T^l_{qi})^*d_i^\dagger c_{ql} \right\},
\label{H_T}
\end{equation}
where $l\in\{ L,R\}$ refers to the left or the right lead, and $\chi^l(t)$ is
the time-dependent switching function of the coupling. The operators $c_{ql}$
and $c_{ql}^\dagger$ annihilate and create an electron in the $l$-lead with a
quantum number $q$ referring both to the continuous momentum $q$ and the subband $n^l$,
see Fig.\ \ref{EB-e-EM-rof} for the corresponding energy spectrum. 
To represent the geometry of the leads and the 
central system, the coupling tensor $T^l_{qi}$ of single-electron states
$|q\rangle$ in the lead $l$ to single-electron states states $|i\rangle$ in the 
system is modeled as a non-local
overlap integral of the corresponding wave functions in the contact
regions of the system, $\Omega_S^l$, and the lead $l$,
$\Omega_l$\cite{Gudmundsson09:113007}
\begin{equation}
      T^l_{iq} = \int_{\Omega_S^l\times \Omega_l} d{\bf r}d{\bf r}'
      \left[\psi^l_q ({\bf r}') \right]^*\psi^S_i({\bf r})\;
      g^l_{iq} ({\bf r},{\bf r'}).
\label{T_aq}
\end{equation}
The function
\begin{equation}
      g^l_{iq} ({\bf r},{\bf r'}) =
                   g_0^l\exp{\left[-\delta_1^l(x-x')^2-\delta_2^l(y-y')^2\right]}
                   \times\exp{\left(\frac{-|E_i-\epsilon^l(q)|}{\Delta_E^l}\right)}
\label{gl}
\end{equation}
with ${\bf r}\in\Omega_\mathrm{S}^l$ and ${\bf r}'\in\Omega_l$
defines the `nonlocal Gaussian overlap' determined by the constants
$\delta_1$ and $\delta_2$, and the affinity of the states in energy $\Delta_E^l$.
The energy spectra of the leads are represented by $\epsilon^l(q)$.

The time-evolution of the total system is determined by the Liouville-von Neumann equation
\begin{equation}
      i\hbar\dot W(t)=[H(t),W(t)],\quad W(t<0)=\rho_\mathrm{L}\rho_\mathrm{R}\rho_\mathrm{S},
\label{L-vN}
\end{equation}
with $W$ the statistical operator of the total system
and $\rho_l$ the equilibrium density operator of the disconnected lead $l\in\{L,R\}$ having
chemical potential $\mu_l$ 
\begin{equation}
      \rho_l=\frac{e^{-\beta (H_l-\mu_l N_l)}}{{\rm Tr}_l \{e^{-\beta(H_l-\mu_l N_l)}\}},
\label{rho_l}
\end{equation}
where $H_l$ is the Hamiltonian of the electrons in lead $l\in\{\mathrm{L,R}\}$ 
and $N_l$ is their number operator.
The Liouville-von Neumann equation (\ref{L-vN})
is projected on the central system of coupled electrons and photons by a partial 
tracing operation with respect to the operators of the leads. 
Defining the reduced density operator (RDO) of the central system
\begin{equation}
      \rho_\mathrm{S}(t)={\rm Tr}_\mathrm{L} {\rm Tr}_\mathrm{R} W(t),
      \quad \rho_\mathrm{S}(0)=\rho_\mathrm{S},
\label{rdo}
\end{equation}
we obtain an integro-differential equation for the RDO, the generalized
master equation (GME)
\begin{align}
      \dot{\rho}_\mathrm{S}(t)=-\frac{i}{\hbar}\left[H_\mathrm{S}, {\rho}_\mathrm{S}(t)\right]
      -\frac{\mathrm{Tr}_\mathrm{LR}}{\hbar^2}
      \left\{\left[ H_\mathrm{T}(t),
      \int_0^t dt'\; \left[ U(t-t')H_\mathrm{T}(t')U^+(t-t'), \qquad\quad{\vphantom{.}}
      \right. \right. \right. \nonumber\\
      \left. \left. \left. 
      U_0(t-t')\rho_\mathrm{S}(t')U_0^+(t-t')\rho_\mathrm{L}\rho_\mathrm{R}\right] \vphantom{\sum}
      \right]\right\} ,
\label{RDO}
\end{align}
where the time evolution operator for the closed systems of Coulomb interacting electrons 
coupled to photons on one hand, and on the other hand noninteracting electrons in the leads is
given by $U(t)=\exp{\{-i(H_\mathrm{e}+H_\mathrm{Coul}+H_\mathrm{EM}+
H_\mathrm{L}+H_\mathrm{R})t/\hbar\}}$, 
without the coupling to the leads $H_\mathrm{T}(t)$.
Here, $H_\mathrm{e}$ is the Hamiltonian for the electrons in the central system,
$H_\mathrm{Coul}$ their mutual Coulomb interaction, $H_\mathrm{EM}$ is the 
Hamiltonian for the photons in the cavity together with their interaction to the 
electrons, and $H_\mathrm{L,R}$
are the Hamiltonian operators for the electrons in the left and right leads.
The time evolution of the closed system of Coulomb interacting electrons 
interacting with the photons is governed by $U_0(t)=\exp{\{-iH_0t/\hbar\}}$.
The GME (\ref{RDO}) is valid in the weak system-leads coupling limit since we
have only retained terms of second order in the coupling Hamiltonian $H_\mathrm{T}$
in its integral kernel. It should though be stressed that we are not approximating
the GME to second order in the coupling as its integral structure effectively 
provides terms of any order, but with a structure reflecting the type of the
integral kernel. 

Commonly, the GME is written in terms of spectral densities for the states in the
system instead of the coupling tensor (\ref{T_aq}). We do not make this transformation
but the spectral densities for the 10 lowest SESs used in our system have been presented
elsewhere \cite{Gudmundsson12:1109.4728} (Here the overall coupling is one quarter of the value
used in Ref.\ \cite{Gudmundsson12:1109.4728}). The spectral density 
is a particularly useful physical concept as it 
demonstrates the spectral broadening of the SESs in the finite system due to the coupling to the leads.

\section{Transport characteristics}
The time-dependent coupling betweeen the leads and the central system is modeled by
the switching functions $\chi^l(t)$.  These functions may be considered input elements
of the transport problem: stepwise functions, periodic, relatively phase-shifted, etc. 
In the following examples the
left and right leads are coupled simultaneously smoothly to the central system
by use of the switching function
\begin{equation}
      \chi^l(t)=\left(1-\frac{2}{e^{\alpha^lt}+1}\right),\quad l\in\{ L,R\}
\label{chi}
\end{equation}
with $\alpha^l = 0.3$ ps$^{-1}$. The temperature of the leads $T=0.5$ K, and the 
overall coupling strength $g_0^la_w^{3/2}=13.3$ meV, is much lower than in our 
earlier calculation \cite{Gudmundsson12:1109.4728}. We choose $\delta_{1,2}^la_w^2=0.4916$ 
meaning that states of a lead and the central system with considerable charge density within a length 
equivalent to $2a_w$ ($a_w\approx 33.5$ nm here) could be well coupled.

The energy of the photon mode is $\hbar\omega =0.4$ meV, and
the electron confinement in the $y$-direction has the energy scale
$\hbar\Omega_0=1.0$ meV. The energy separation of the lowest states
for the motion in the $x$-direction is lower, $\sim 0.2$ meV.
The Coulomb interaction has a characteristic energy scale 
$0.5e^2/(\kappa a_w)\approx 1.7$ meV. The external magnetic field is low  
enough so that only the highest lying SESs show any effects of the 
Lorentz force. We expect thus the transport properties of the
system to be anisotropic at a low energy scale with respect to polarization of the 
photon field along ($x$-direction) or perpendicular ($y$-direction) to the transport. 

\begin{figure}[htbq]
            \includegraphics[width=0.49\textwidth,angle=0,bb=50 50 396 395,clip]{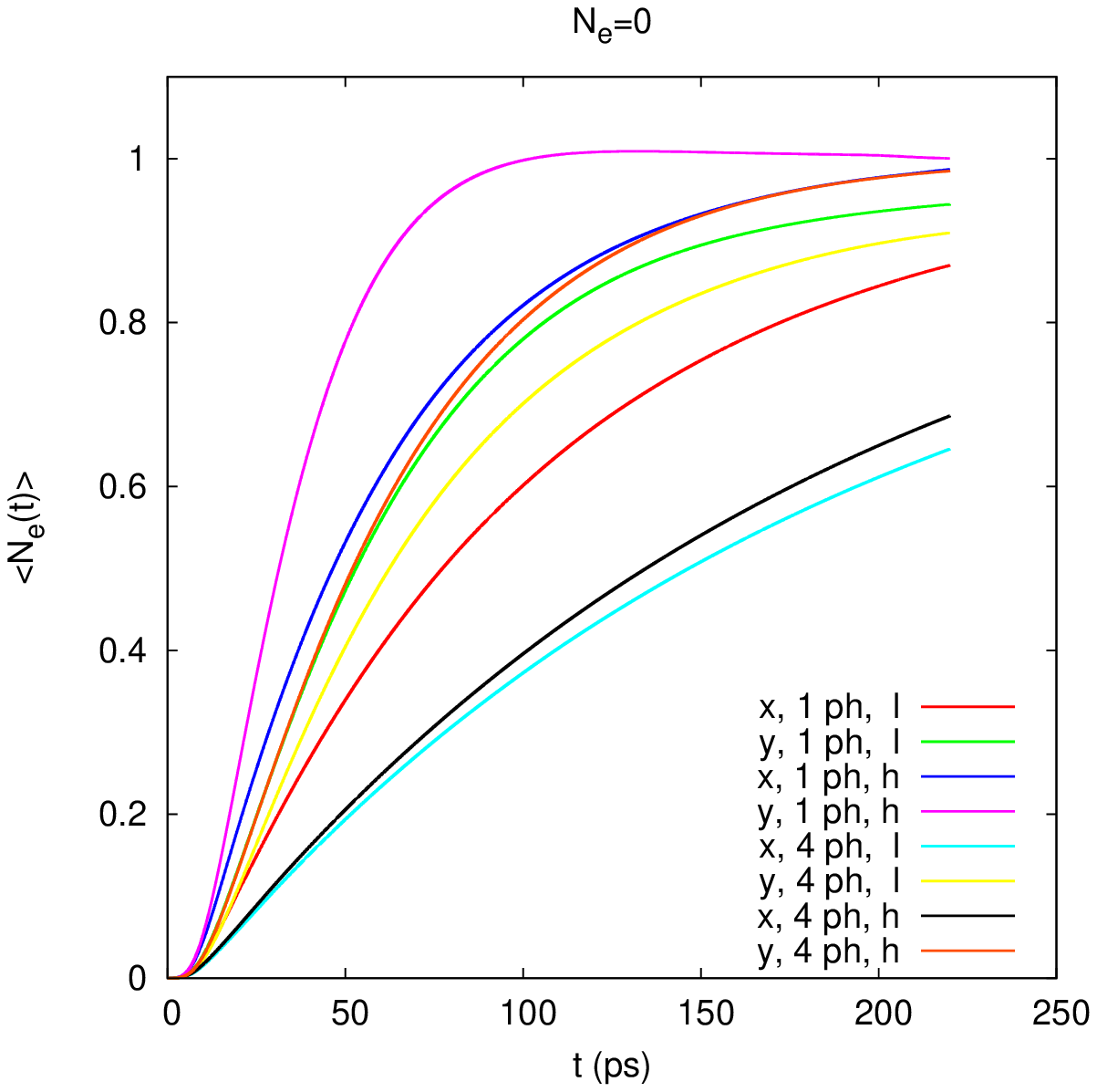}
            \includegraphics[width=0.49\textwidth,angle=0,bb= 1  1 348 344,clip]{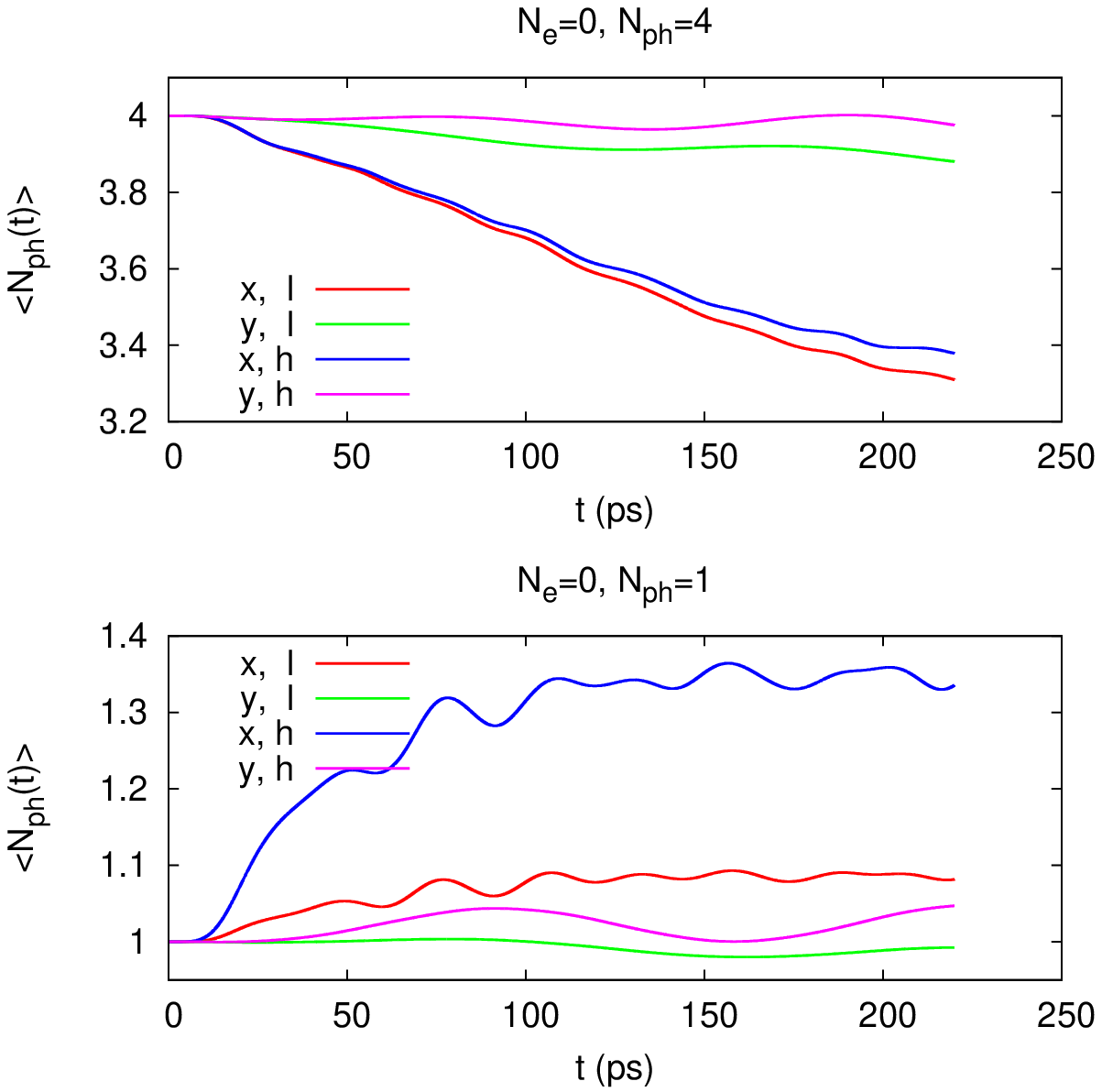}
      \caption{The total mean number of electrons $\langle N_\mathrm{e}(t)\rangle$ (left panel),
               and the total mean photon number $\langle N_\mathrm{ph}(t)\rangle$ (right panels)
               as a function of time, polarization of the photon field ($x$ or $y$)
               and bias window low (l: $\mu_L=2.0$ and $\mu_R=1.4$ meV) or 
               high (h: $\mu_L=3.0$ and $\mu_R=2.5$ meV).  
               $B=0.1$ T, $g^\mathrm{EM}=0.1$ meV, $\hbar\Omega_0=1.0$ meV, $L_x=300$ nm, 
               $\hbar\omega = 0.4$ meV,
               $\Delta^l_E=0.25$ meV, $g_0^la_w^{3/2}=13.3$ meV,
               $\delta_{1,2}^la_w^2=0.4916$, $m^*=0.067m_e$, and $\kappa =12.4$.}
      \label{Ne-Nph-heild}
\end{figure}
In the present calculations we start with a central system empty of electrons, but with
one or four photons in the cavity. 
For $N_\mathrm{SES}=18$ the state with no electron, but one photon, is $|\breve{2})$
for both polarizations. The state containing 4 photons and no electrons is $|\breve{14})$
for the $x$-polarization and $|\breve{15})$ in the case of the $y$-polarization.
The time evolution of the total mean electron number
$\langle N_\mathrm{e}(t)\rangle$ is seen in the left panel of Fig.\ \ref{Ne-Nph-heild} 
and the total mean number of photons $\langle N_\mathrm{ph}(t)\rangle$ is displayed in the lower 
right panel for the case of initially one photon and in the upper panel for initially 4 photons.   
The charging of the system is in most cases fastest for the higher bias window as the active states of
the central system are well coupled to states in the leads that carry a large current in
this energy range. The presence of photons in the system sharply diminishes the charging
speed, especially for their polarization along the transport direction ($x$-direction).
The time evolution of the total mean number of photons in the system does not give much
insight into what is happening in the system, but it is though clear that it varies more
in case of the $x$-polarization.  

\begin{figure}[htbq]
      \begin{center}
            \includegraphics[width=0.48\textwidth,angle=0]{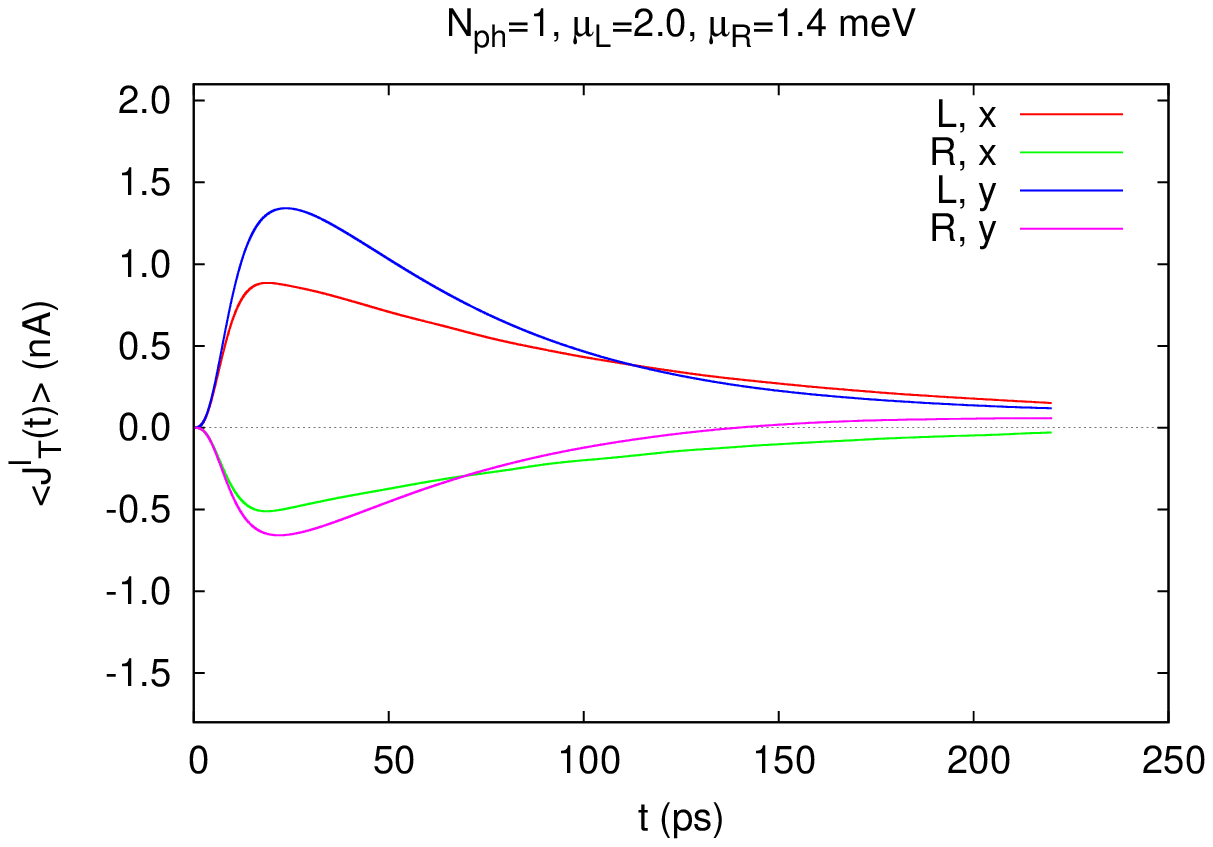}
            \includegraphics[width=0.48\textwidth,angle=0]{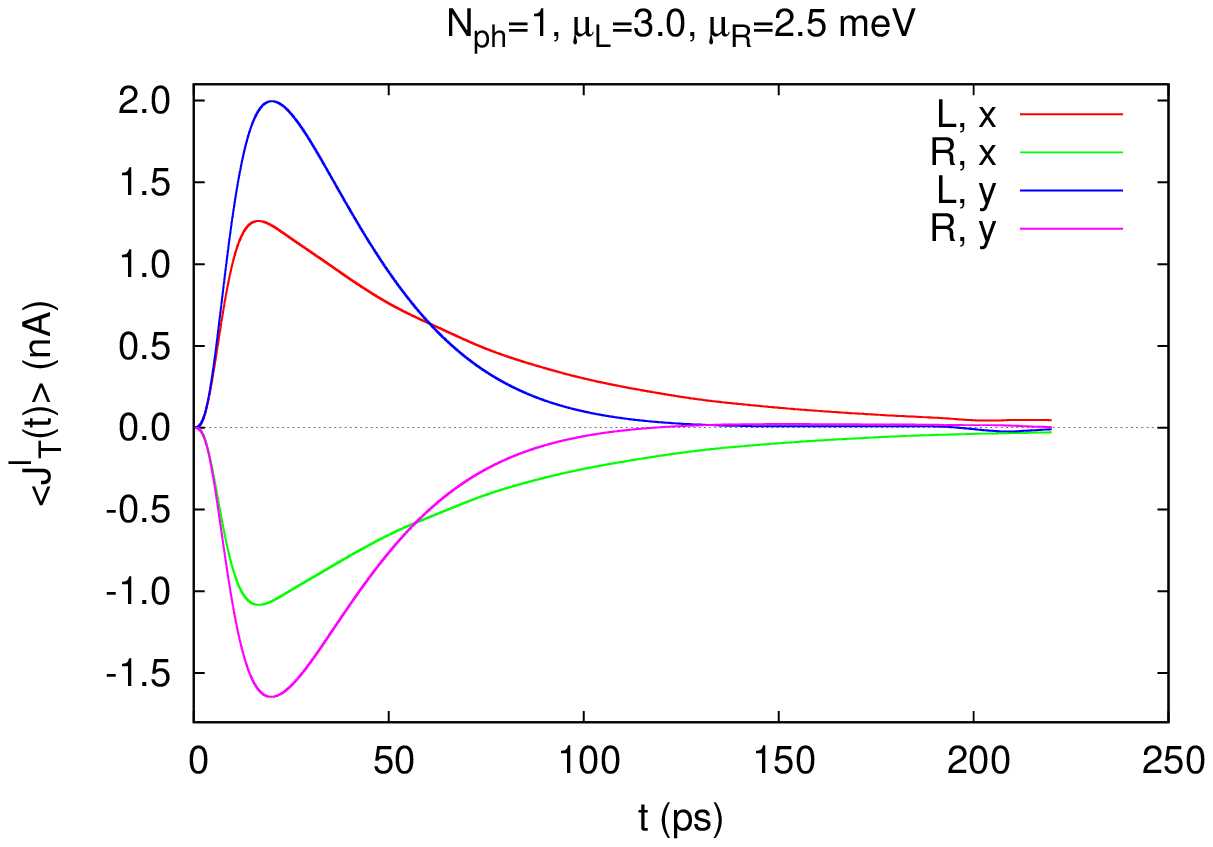}\\
            \includegraphics[width=0.48\textwidth,angle=0]{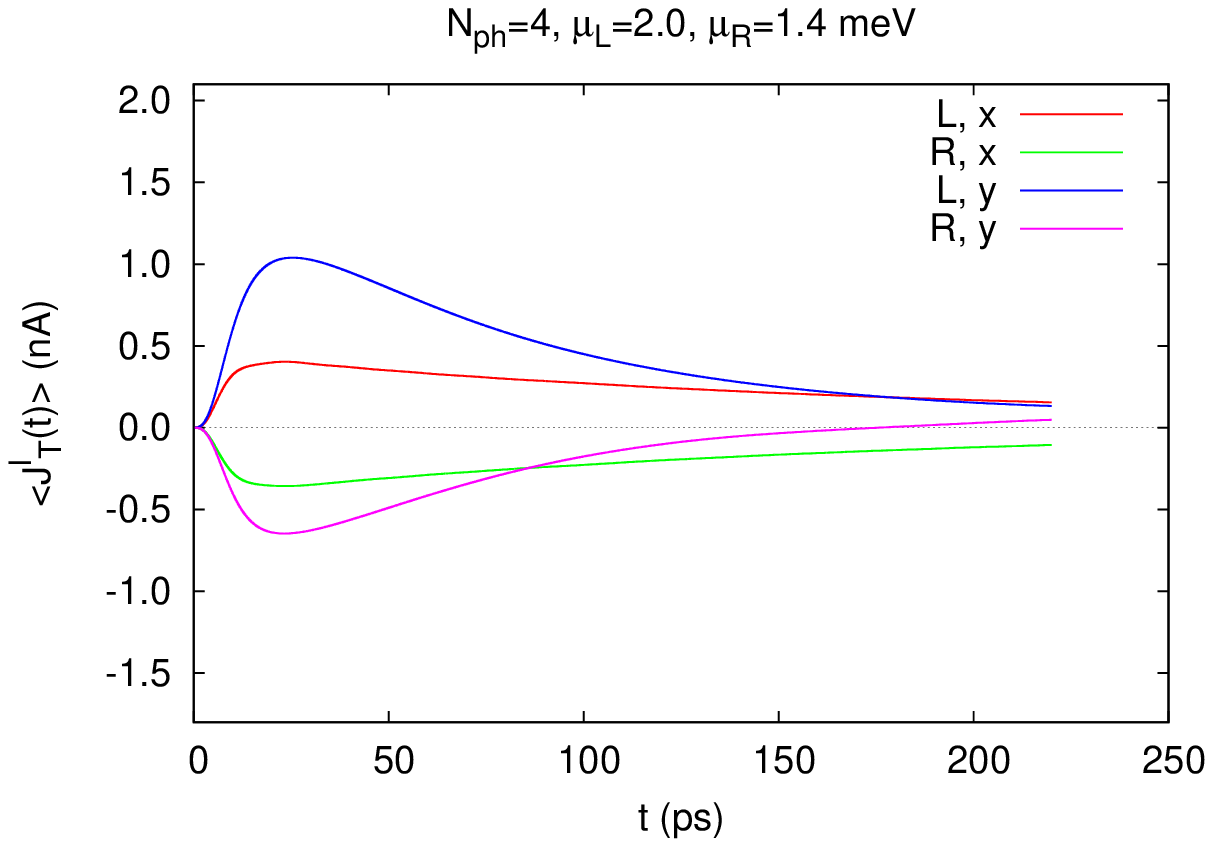}
            \includegraphics[width=0.48\textwidth,angle=0]{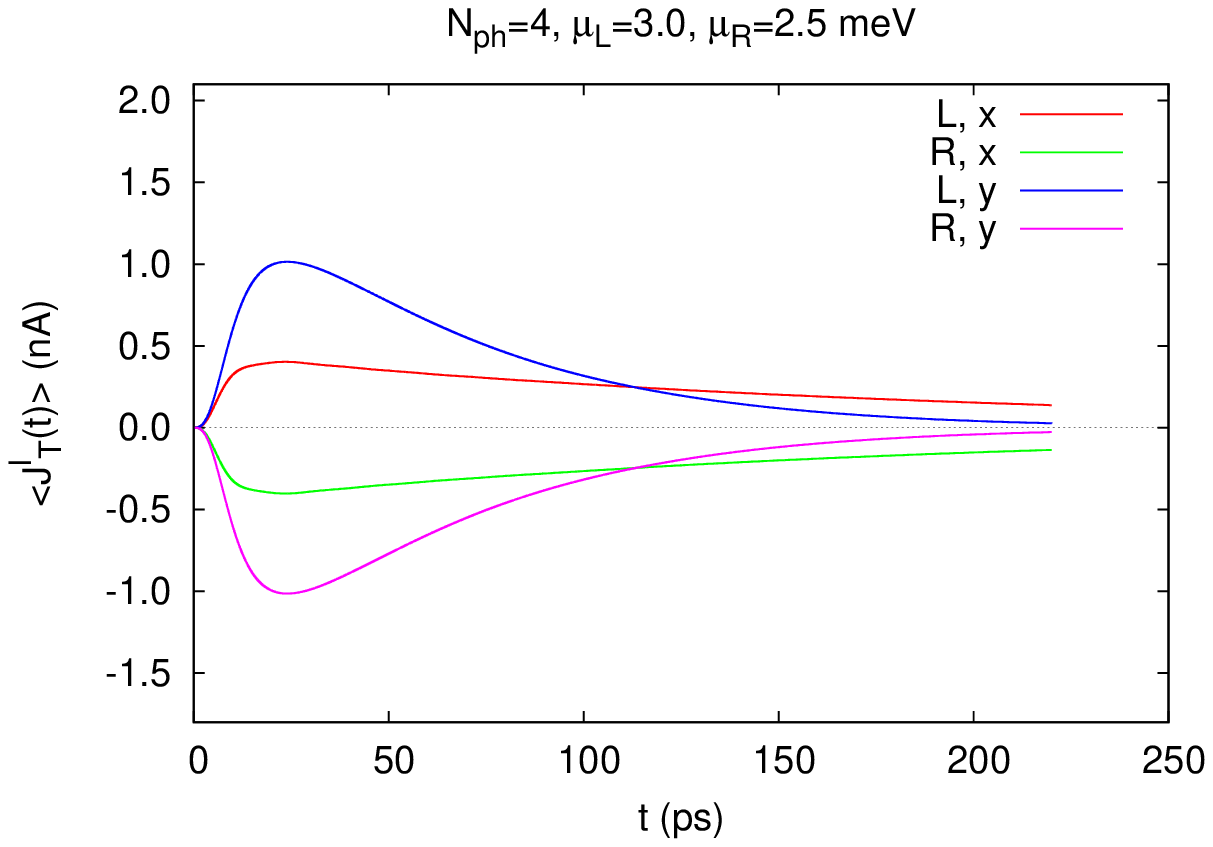}
      \end{center}
      \caption{The total current from the left lead (L), and into the right lead (R) 
               as a function of time for $g^\mathrm{EM}=0.1$ meV
               and polarization of the photon field ($x$ or $y$). Initially, at $t=0$ there is one photon 
               in the cavity (top panel), or 4 photons (bottom panel). The bias window is
               low ($\mu_L=2.0$ and $\mu_R=1.4$ meV) for the left panels and high 
               ($\mu_L=3.0$ and $\mu_R=2.5$ meV) for the right panels.  
               $B=0.1$ T, $\hbar\Omega_0=1.0$ meV, $L_x=300$ nm, $\hbar\omega = 0.4$ meV,
               $\mu_L=2.0$ meV, $\mu_R=1.4$ meV, $\Delta^l_E=0.25$ meV, $g_0^la_w^{3/2}=13.3$ meV,
               $\delta_{1,2}^la_w^2=0.4916$,
               $m^*=0.067m_e$, and $\kappa =12.4$.}
      \label{J-myndir}
\end{figure}

Very similar information can be read from the graphs of the total currents 
in the left (L) or the right (R) leads shown in Fig.\ \ref{J-myndir}.
The negative values for the current into the right lead indicate a flow from the lead
to the central system. 
A note of caution here is that we are concentrating on the charging time-regime here and we are
not trying to reach a possible steady state, but clearly we are most likely in a Coulomb blocking range.
It is our experience that a considerable probability for two electrons in the system 
is only seen after several two-electron states are in or under the bias window. The reason for this
is most likely the different coupling between states in the leads and the system, how far the system
is from equilibrium and, and how the coupling strength influences the rate of occupation of 
various dynamically correlated states. We have verified that a still higher bias leads to a
nonvanishing steady state current.
Here, we always have at least one photon initially in the cavity and we notice that the charging rate
and the currents are mostly higher for the $y$-polarization. The time-scale for the charging in the 
$x$-polarization gets very long as the photon number is increased from 1 to 4. This is not a total
surprise since the energy of the photons is closer to characteristic excitation energies for a
motion in the $x$-direction.  

\begin{figure}[htbq]
      \begin{center}
            \includegraphics[width=0.40\textwidth,angle=0,bb=20 18 330 212,clip]{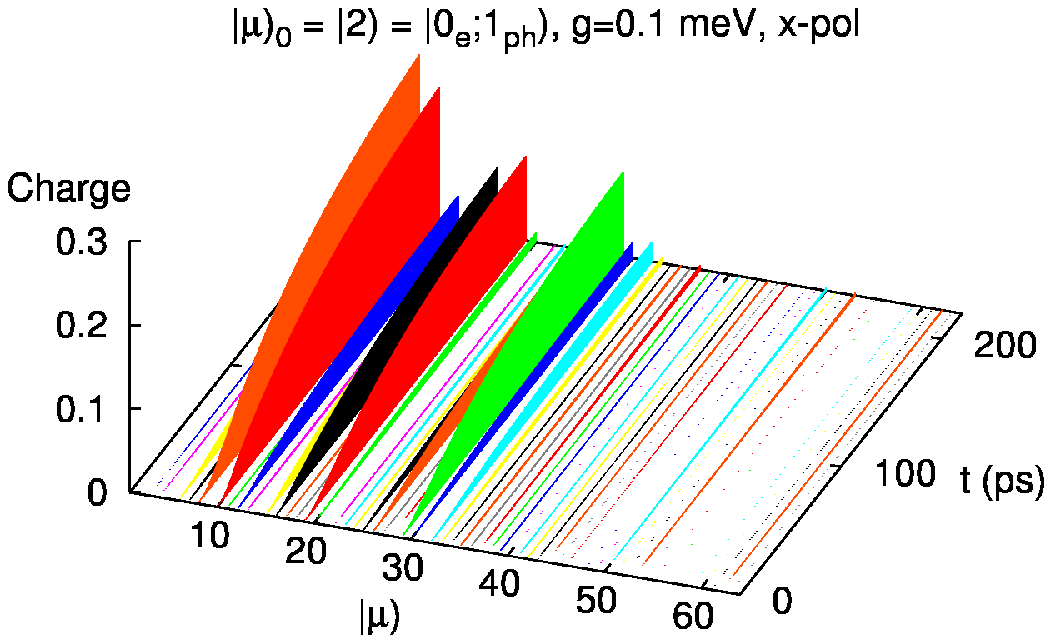}
            \includegraphics[width=0.40\textwidth,angle=0,bb=20 18 330 212,clip]{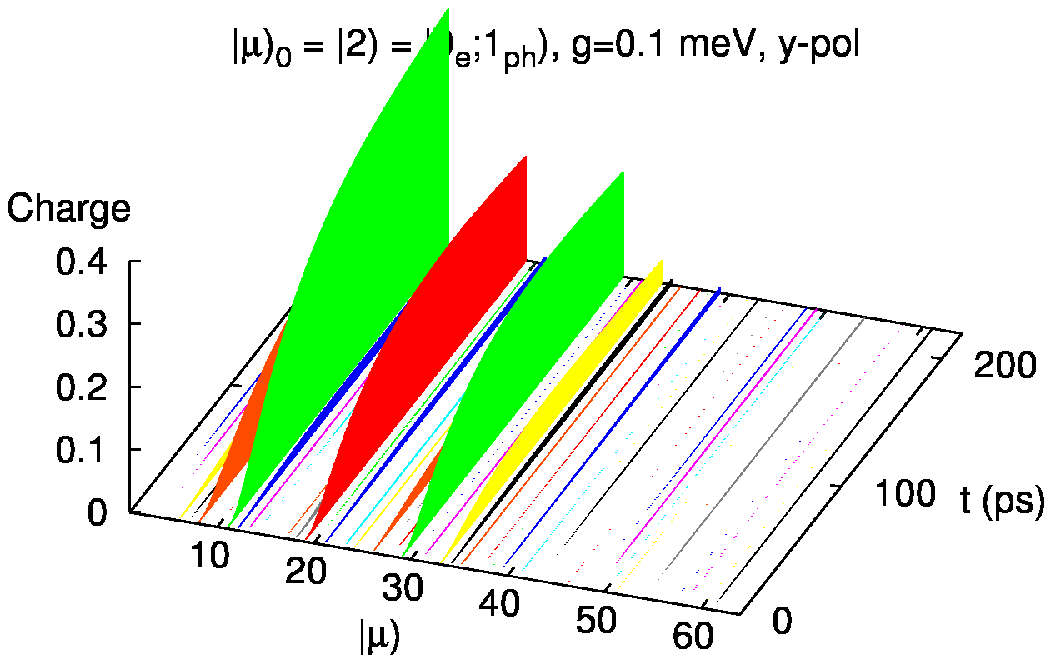}\\
            \includegraphics[width=0.40\textwidth,angle=0,bb=20 18 330 212,clip]{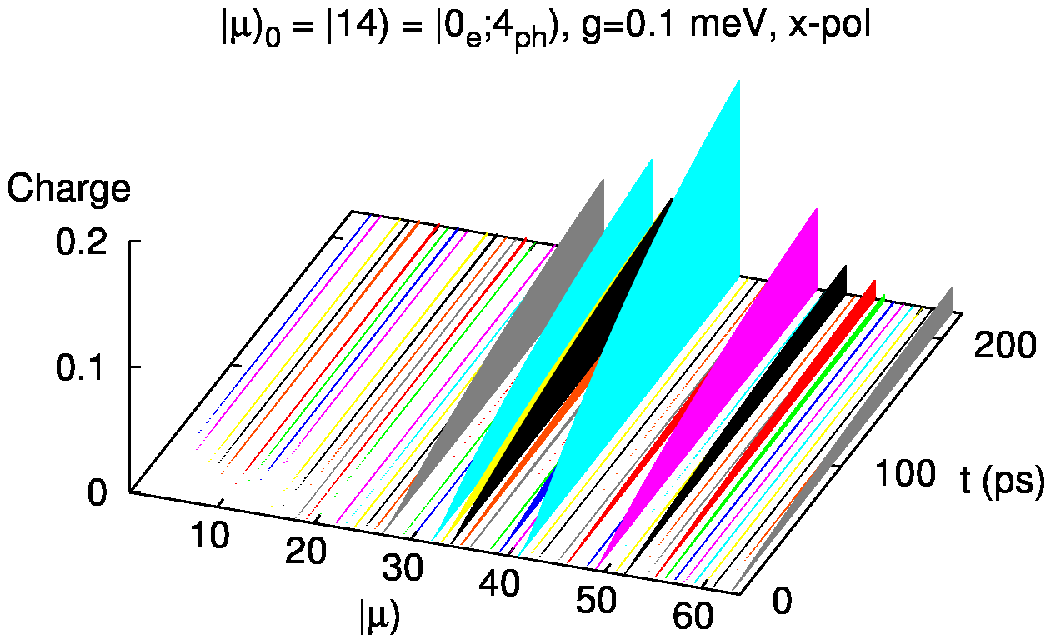}
            \includegraphics[width=0.40\textwidth,angle=0,bb=20 18 330 212,clip]{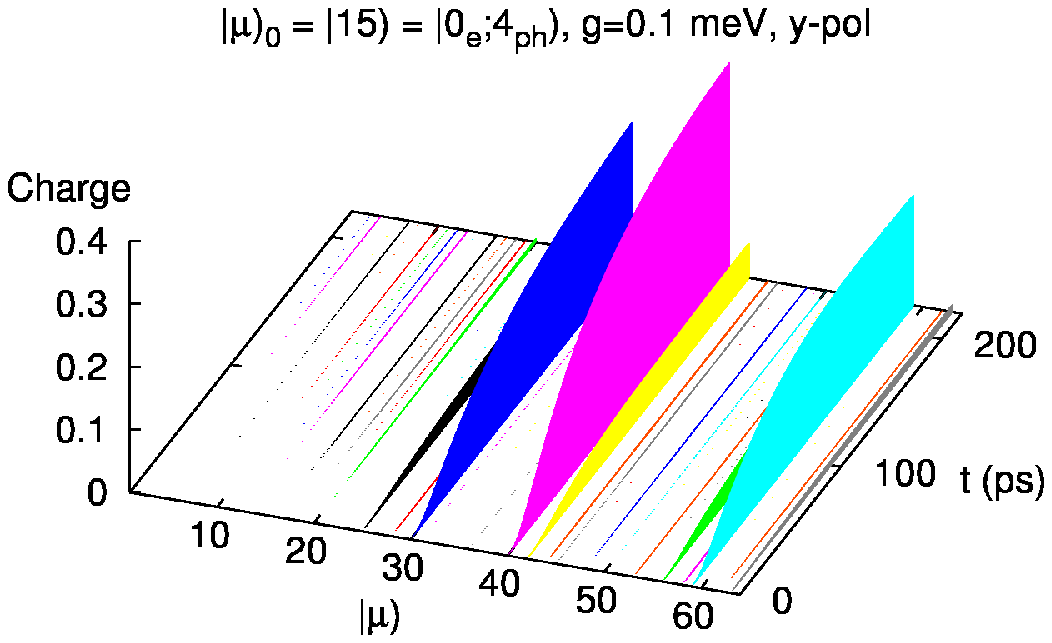}
      \end{center}
      \caption{The mean number of electrons $\langle N_\mathrm{e}(t)\rangle$
               in a MBS $|\breve{\mu})$ 
               for a low bias window ($\mu_L=2.0$ and $\mu_R=1.4$ meV)             
               for $x$-polarization (left) and $y$-polarization (right)
               as a function of time. Initially, at $t=0$, there is one photon 
               in the cavity (top panels), or 4 photons (bottom panels).
               The initial number of electrons is zero in all cases.
               $B=0.1$ T, $g^\mathrm{EM}=0.10$ meV, $\hbar\omega = 0.4$ meV,
               $\hbar\Omega_0=1.0$ meV, $\Delta^l_E=0.25$ meV, $g_0^la_w^{3/2}=13.3$ meV,
               $\delta_{1,2}^la_w^2=0.4916$, $L_x=300$ nm, $m^*=0.067m_e$, and $\kappa =12.4$.}
      \label{Charge-l-myndir}
\end{figure}

A detailed view of the charging processes can be obtained by observing the time-dependent 
probabilities for occupation of the available many-body states (MBS) by electrons or photons. 
In Fig.\ \ref{Charge-l-myndir}
we see the mean charge in the MBS for the transport through the lower bias window 
$\mu_L=2.0$ and $\mu_R=1.4$ meV.
The MBS $|\breve{\mu})$ are numbered according to increasing energy. We see clearly that the 
system is not close to equilibrium and the charge is ``scattered'' to more states for the 
$x$-polarization than the $y$-polarization. The presence of photons has large effects on the 
electrons in the system. 

\begin{figure}[htbq]
      \begin{center}
            \includegraphics[width=0.40\textwidth,angle=0,bb=20 18 330 212,clip]{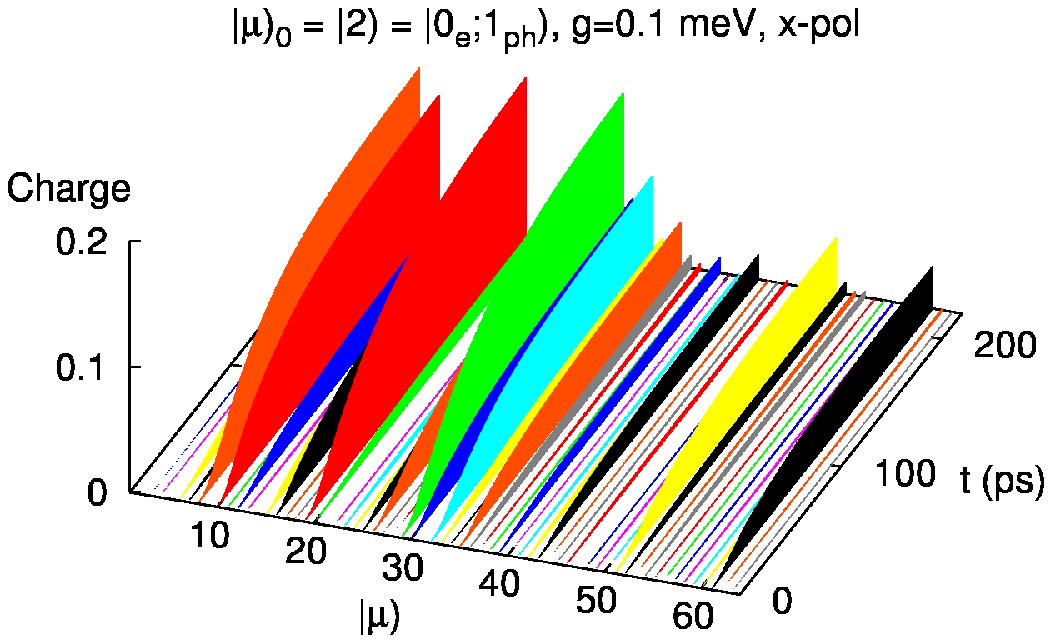}
            \includegraphics[width=0.40\textwidth,angle=0,bb=20 18 330 212,clip]{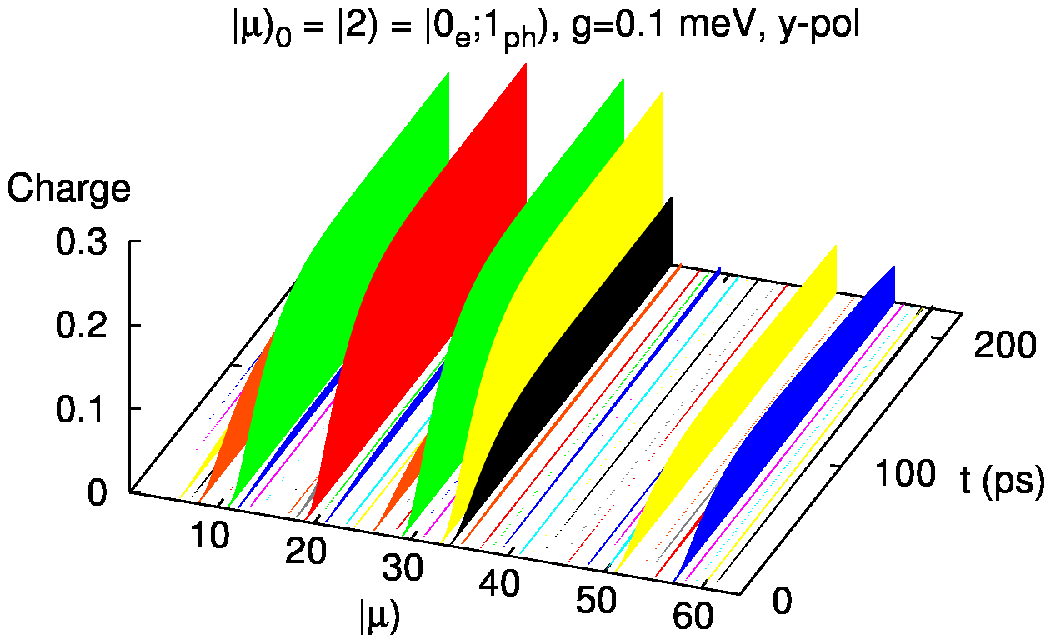}\\
            \includegraphics[width=0.40\textwidth,angle=0,bb=20 18 330 212,clip]{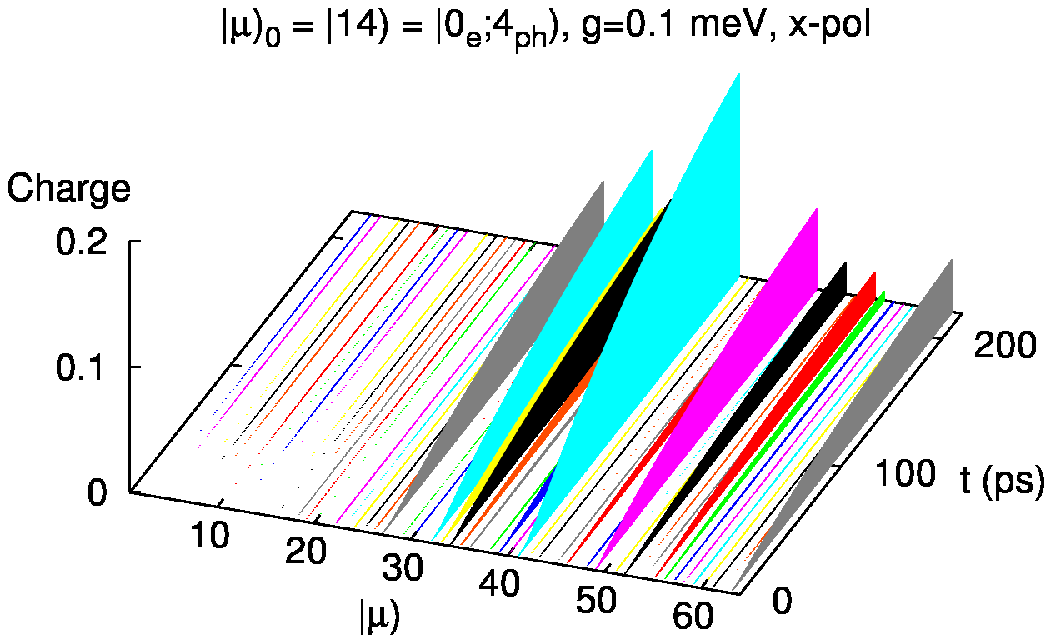}
            \includegraphics[width=0.40\textwidth,angle=0,bb=20 18 330 212,clip]{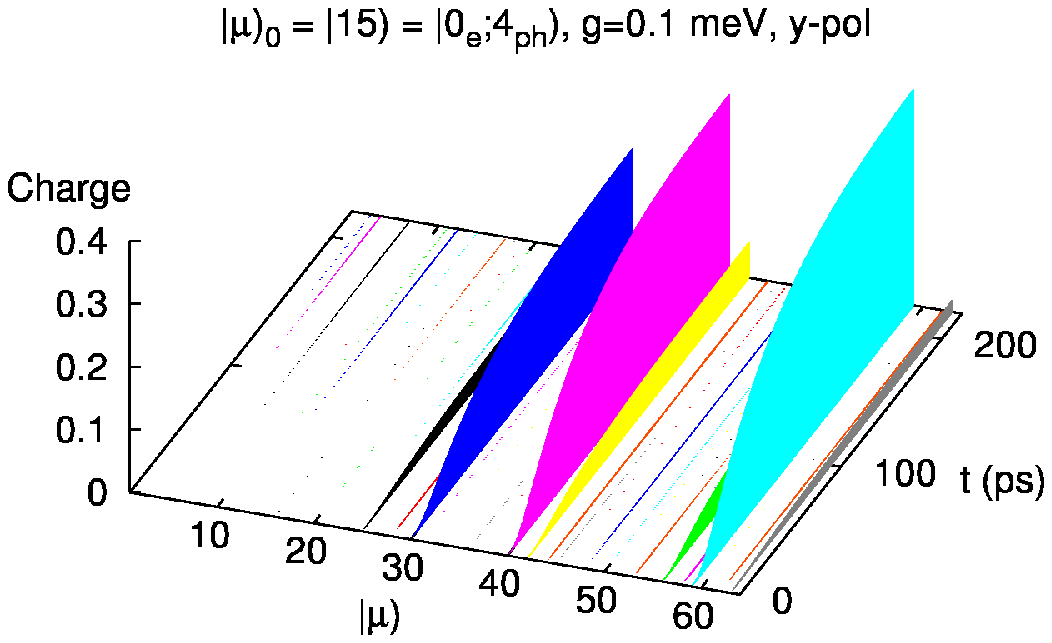}
      \end{center}
      \caption{The mean number of electrons $\langle N_\mathrm{e}(t)\rangle$
               in a MBS $|\breve{\mu})$ 
               for a high bias window ($\mu_L=3.0$ and $\mu_R=2.5$ meV)             
               for $x$-polarization (left) and $y$-polarization (right)
               as a function of time. Initially, at $t=0$, there is one photon 
               in the cavity (top panels), or 4 photons (bottom panels).
               The initial number of electrons is zero in all cases.
               $B=0.1$ T, $g^\mathrm{EM}=0.10$ meV, $\hbar\omega = 0.4$ meV,
               $\hbar\Omega_0=1.0$ meV, $\Delta^l_E=0.25$ meV, $g_0^la_w^{3/2}=13.3$ meV,
               $\delta_{1,2}^la_w^2=0.4916$, $L_x=300$ nm, $m^*=0.067m_e$, and $\kappa =12.4$.}
      \label{Charge-h-myndir}
\end{figure}

Very similar story can be said about the results for the higher bias window, 
$\mu_L=3.0$ and $\mu_R=2.5$ meV displayed in Fig.\ \ref{Charge-h-myndir}.
Especially interesting is to see that the results for both polarizations are almost identical
for the higher and lower bias window for the case of 4 photons initially in the central 
system. The effects of the bias window are washed out by the strong interaction with the
quantized electromagnetic field in the cavity. We have though to bring in a note of caution 
here and admit that this effect should be studied further using a larger MB basis for the 
RDO. 

\begin{figure}[htbq]
      \begin{center}
            \includegraphics[width=0.40\textwidth,angle=0,bb=20 18 330 212,clip]{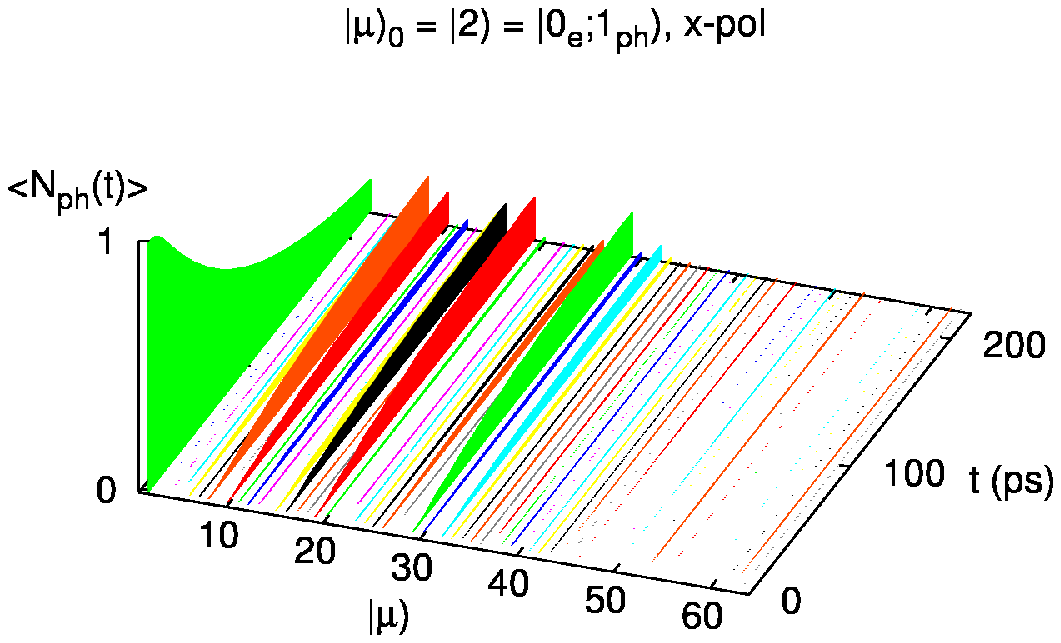}
            \includegraphics[width=0.40\textwidth,angle=0,bb=20 18 330 212,clip]{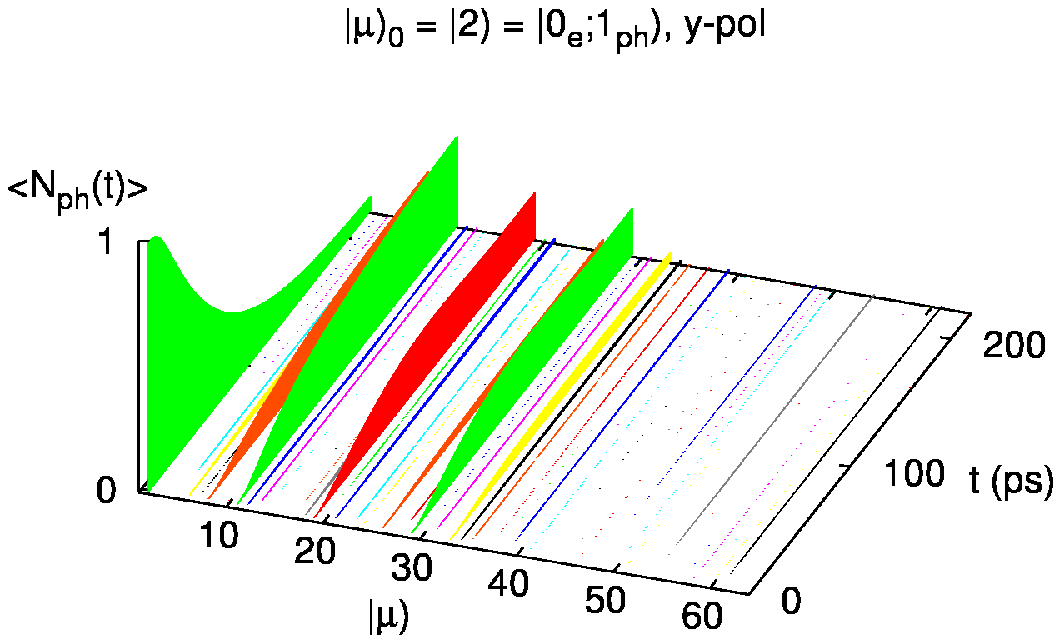}\\
            \includegraphics[width=0.40\textwidth,angle=0,bb=20 18 330 212,clip]{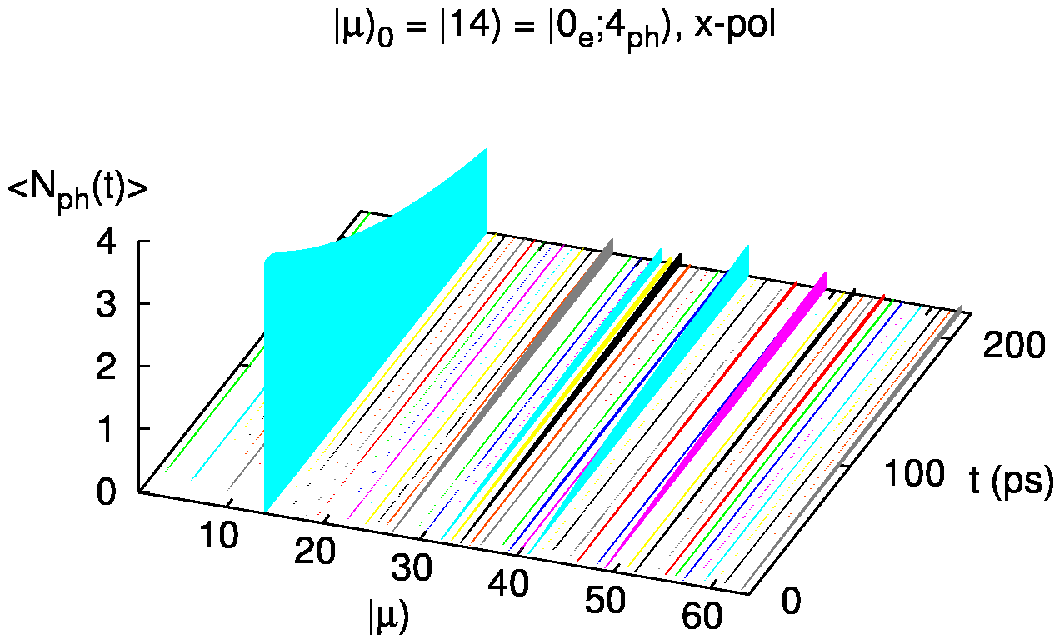}
            \includegraphics[width=0.40\textwidth,angle=0,bb=20 18 330 212,clip]{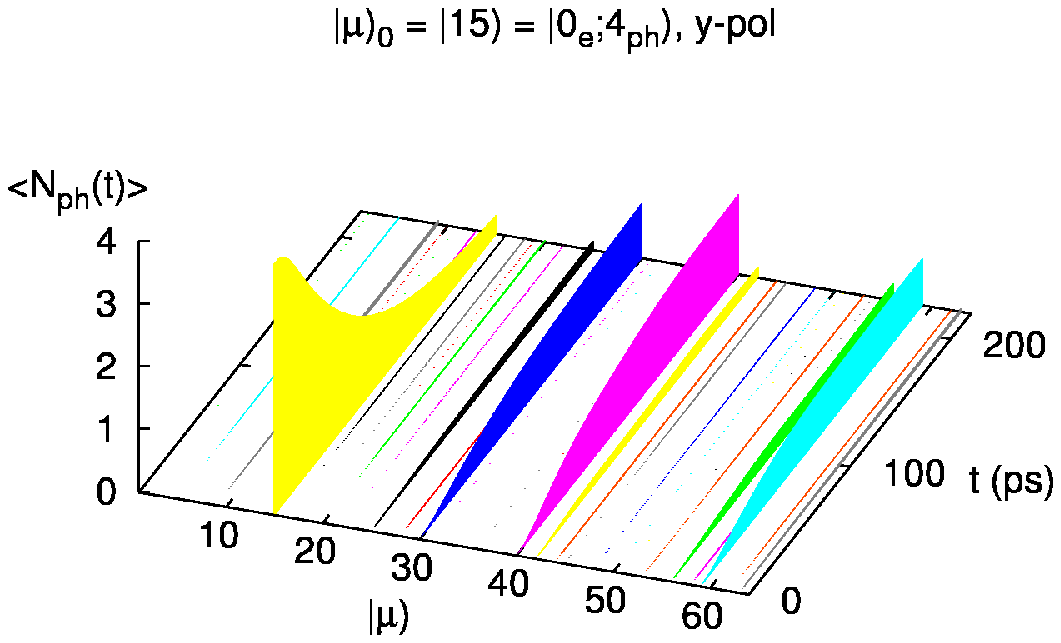}
      \end{center}
      \caption{The mean number of photons $\langle N_\mathrm{ph}(t)\rangle$ 
               in a MBS $|\breve{\mu})$ 
               for a low bias window ($\mu_L=2.0$ and $\mu_R=1.4$ meV)
               for $x$-polarization (left) and $y$-polarization (right)
               as a function of time. Initially, at $t=0$, there is one  photon 
               in the cavity (top panels), or 4 photons (bottom panels).
               $B=0.1$ T, $g^\mathrm{EM}=0.1$ meV, $\hbar\omega = 0.4$ meV,
               $\hbar\Omega_0=1.0$ meV, $\Delta^l_E=0.25$ meV, $g_0^la_w^{3/2}=13.3$ meV,
               $\delta_{1,2}^la_w^2=0.4916$, $L_x=300$ nm, $m^*=0.067m_e$, and $\kappa =12.4$.}
      \label{Nph-l-myndir}
\end{figure}

\begin{figure}[htbq]
      \begin{center}
            \includegraphics[width=0.40\textwidth,angle=0,bb=20 18 330 212,clip]{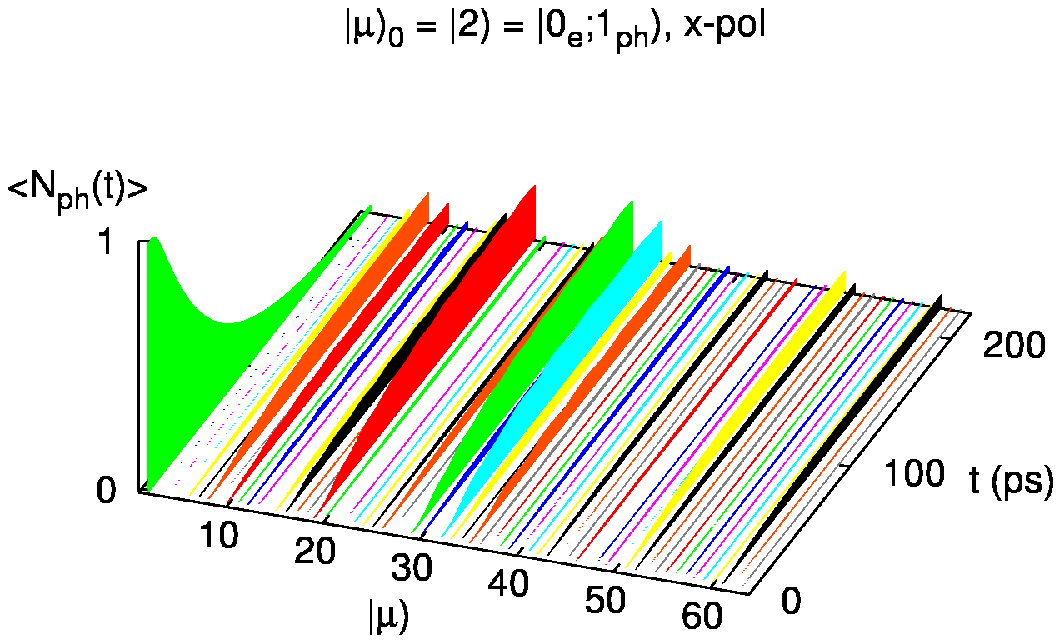}
            \includegraphics[width=0.40\textwidth,angle=0,bb=20 18 330 212,clip]{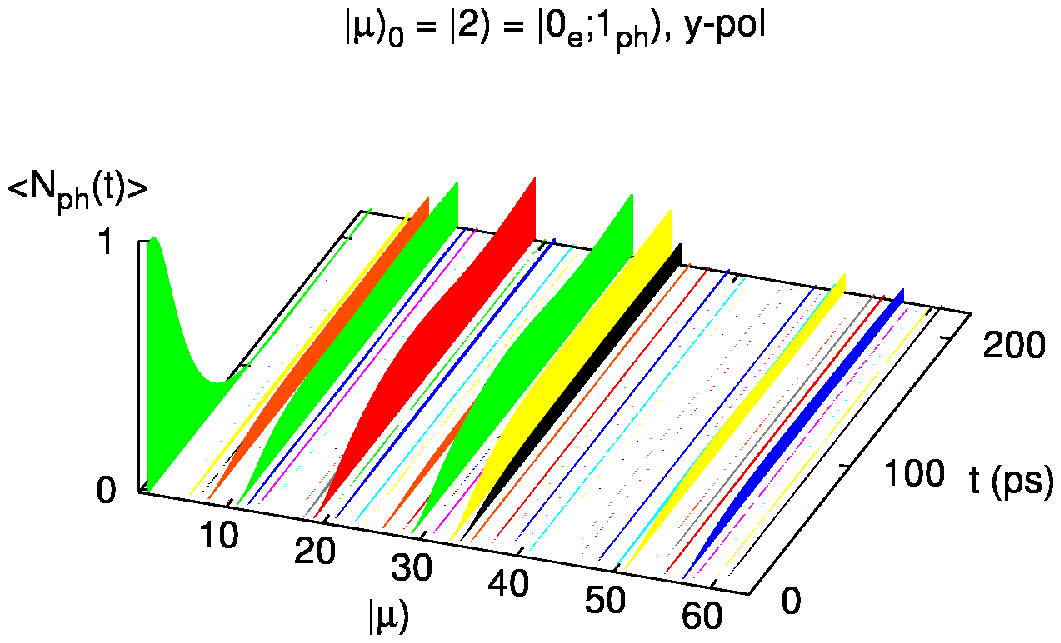}\\
            \includegraphics[width=0.40\textwidth,angle=0,bb=20 18 330 212,clip]{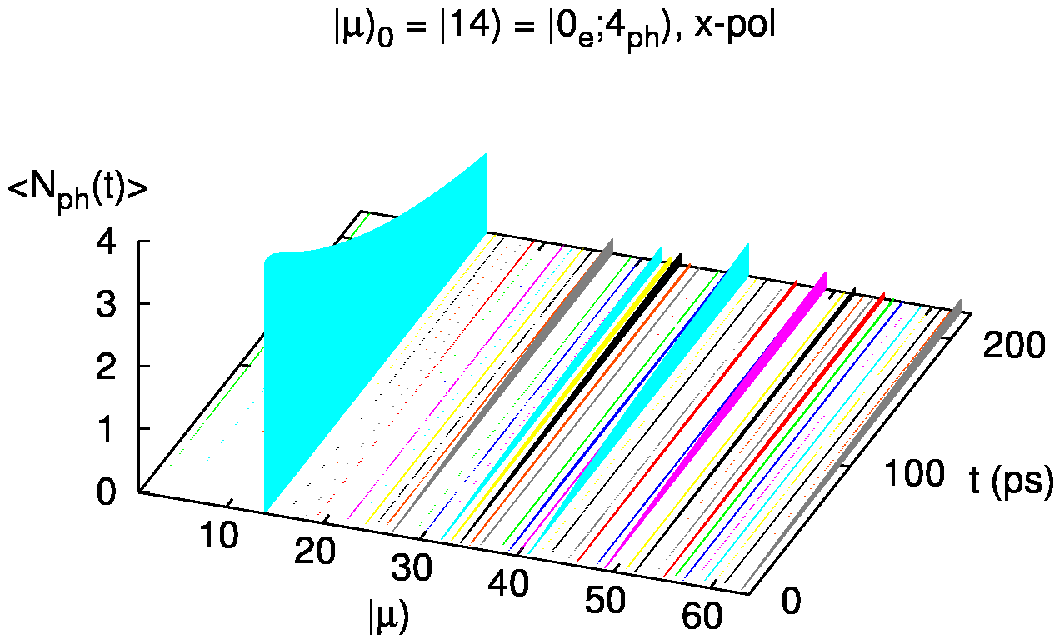}
            \includegraphics[width=0.40\textwidth,angle=0,bb=20 18 330 212,clip]{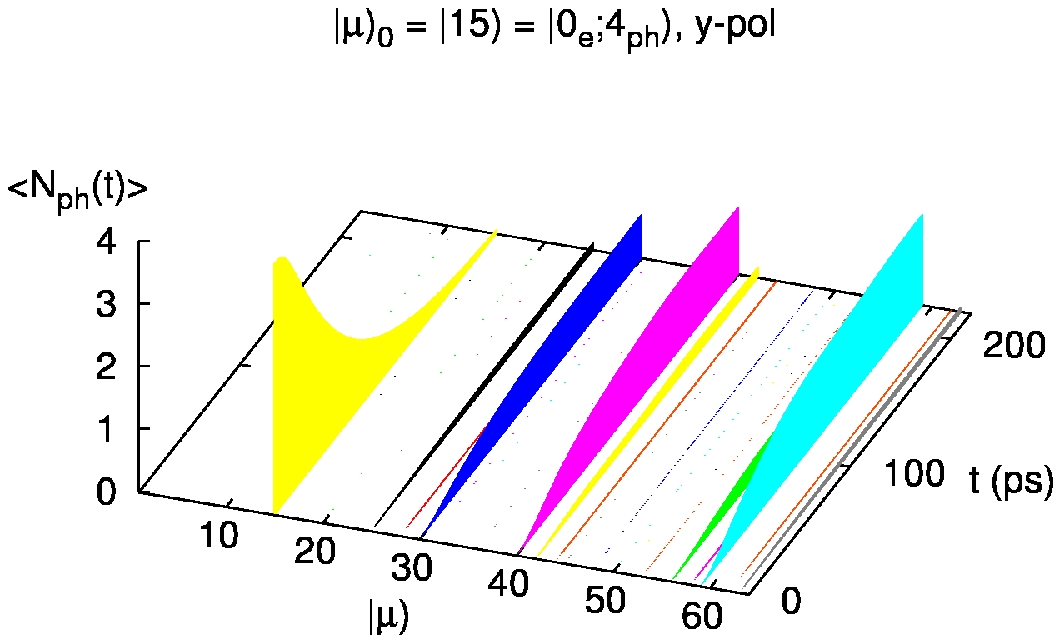}
      \end{center}
      \caption{The mean number of photons $\langle N_\mathrm{ph}(t)\rangle$ 
               in a MBS $|\breve{\mu})$ 
               for a high bias window ($\mu_L=3.0$ and $\mu_R=2.5$ meV)
               for $x$-polarization (left) and $y$-polarization (right)
               as a function of time. Initially, at $t=0$, there is one  photon 
               in the cavity (top panels), or 4 photons (bottom panels).
               $B=0.1$ T, $g^\mathrm{EM}=0.1$ meV, $\hbar\omega = 0.4$ meV,
               $\hbar\Omega_0=1.0$ meV, $\Delta^l_E=0.25$ meV, $g_0^la_w^{3/2}=13.3$ meV,
               $\delta_{1,2}^la_w^2=0.4916$, $L_x=300$ nm, $m^*=0.067m_e$, and $\kappa =12.4$.}
      \label{Nph-h-myndir}
\end{figure}

A complete picture can not be reached without exploring the time-evolution of the 
mean photon number per MBS presented in Fig.\ \ref{Nph-l-myndir} for the lower
bias window and in \ref{Nph-h-myndir} for the higher bias window.
Again we notice that the state of the system, the distribution of the photon component into various
MBS, is almost independent of the the bias window in the case of 4 photons initially in the cavity. 

But, here another very important fact about the system evolution becomes evident. 
The occupation of the initial photon state seems to vary much faster with time than the
total mean number of photons shown in Fig.\ \ref{Ne-Nph-heild}. The slow decay of the charging
current in Fig.\ \ref{J-myndir} for the system with 4 photons initially might also indicate that 
radiation processes here are slow. Why then do we have a fast redistribution of the photon component
between the available MBS initially, during the switch-on process? 
The resolution of this dilemma comes from remembering the 
structure of the interaction terms. Part of the interaction is an integral over the term 
$\mathbf{j}\cdot\mathbf{A}$. We have a vector field initially present in the system.
The high current into the system during the transient phase, when the contacts between the
leads and the central system are switched on, creates a strong interaction that can 
``scatter'' the electrons and the photons to different MBS, few or many, depending on
selections rules and resonances. This initial rushing of electrons from both ends of the finite
quantum wire is a longitudinal (irrotational) current that enhances the coupling to the photon field.      
This effect can be seen very well by comparing the photon distribution into the MBS
at the low bias window and for the case of only one photon initially present in the cavity. 

In case of the $x$-polarization we see that the system seems to develop a higher
``impedance'' acting against its charging as the number of photons is increased,
and at the same time the electrons that enter the system are scattered to a wide
range of MBS. The ``impedance'' has to be understood from the fact that the 
photon field is polarizing the electron density inside the system along the 
$x$-direction. Effectively, the states may gain coupling strength to the leads,
but that works in both ways. The electrons enter the system easily and leave it
equally easily at the same end before they can be transported through it.
The charging of the system is counteracted by an effective scattering of a localized
photon field. A phenomena that should be similar for a system with localized 
phonons. 

\section{Summary}    
In this publication we have shown how we have successfully been able to
implement a scheme which we wish to call: ``A stepwise introduction of complexity 
to a model description and a careful counteracting stepwise truncation of the 
ensuing many-body space'' to describe a time-dependent transport of Coulomb
interacting electrons through a photon cavity. We have used this approach to
demonstrate how the geometry of a particular system leaves its fingerprints 
on its transport properties. We guarantee geometrical dependence by using a 
phenomenological description of the system-lead coupling based on a nonlocal 
overlap of the single-electron states in the leads and the system in the contact area,
and by using a large basis of SESs in order to build our MBS. The Coulomb interaction
is implemented by an ``exact numerical diagonalization'', and the coupling to
the quantized electromagnetic cavity field of a single frequency is carried 
out using both the paramagnetic and the diamagnetic part of the charge current
density. The coupling of the electrons and the photons is also treated by
an exact numerical diagonalization without resorting to the rotating
wave approximation. This approach is thus applicable for the modeling of
circuit-QED elements in the strong coupling limit.

We have deployed the GME method to include memory effects without a Markov
approximation. Moreover, the GME approach is utilized to describe the coupling 
of the electron-photon system to the external leads.
At this moment in the development of the model we ignore the spin variable of
freedom, but it can easily be included. The inclusion of the spin is 
(more or less) straight forward, but it requires a double amount of computer memory
and an increased number of MBSs in the GME solver. This is work in progress now.    

We concentrate our investigations on the charging regime of the system and
find that it is strongly influenced by the presence of photons in the cavity, 
their polarization and the geometry of the system. We consider these results
presented here as the mere initial steps into the exploration of a fascinating
regime of circuit-QED elements. 

For the numerical implementation we rely heavily on parallel processing, but we foresee
further refinement in the truncation schemes for the many-body spaces and in the
parallelization that will allow us to describe systems of increased complexity.

%----------------------------

\begin{acknowledgments}
      The authors acknowledge financial support from the Icelandic Research
      and Instruments Funds,
      the Research Fund of the University of Iceland, the
      National Science Council of Taiwan under contract
      No.\ NSC100-2112-M-239-001-MY3. HSG acknowledges support from the National Science Council, 
      Taiwan, under Grant No.\ 100-2112-M-002-003-MY3, 
      support from the Frontier and Innovative Research Program of the National Taiwan University 
      under Grants No.\ 10R80911 and No. 10R80911-2, and support from the
      focus group program of the National Center for Theoretical Sciences, Taiwan.
\end{acknowledgments}

% Use this code if you wish to generate your bibliography with BibTeX;
% please replace first the string "demo" below with the name(s) of
% the BibTeX data base(s) you want to use.
% The resulting bibliography-output (the contents of the .bbl file)
% must be pasted into this file before submission.
% 
\bibliographystyle{fdp}

\begin{thebibliography}{[10]}

\bibitem{Niemczyk10:772}% article
 \textsc{T.~Niemczyk},  \textsc{F.~Deppe},  \textsc{H.~Huebl},  \textsc{E.\,P.
  Menzel},  \textsc{F.~Hocke},  \textsc{M.\,J. Schwarz},  \textsc{J.\,J.
  Garcia-Ripoll},  \textsc{D.~Zueco},  \textsc{T.~H{\"u}mmer},
  \textsc{E.~Solano},  \textsc{A.~Marx},  and  \textsc{R.~Gross}\iffalse
  Circuit quantum electrodynamics in the ultrastrong-coupling regime\fi,
 \jr{Nature Physics} \textbf{6}, 772 (2010).


\bibitem{Frey11:01}% article
 \textsc{T.~Frey},  \textsc{P.\,J. Leek},  \textsc{M.~Beck},
  \textsc{A.~Blais},  \textsc{T.~Ihn},  \textsc{K.~Ensslin},  and
  \textsc{A.~Wallraff}\iffalse Dipole coupling of a double quantum dot to a
  microwave resonator\fi,
 \jr{arXiv:1108.5378} (2011).


\bibitem{Delbecq11:01}% article
 \textsc{M.~Delbecq},  \textsc{V.~Schmitt},  \textsc{F.~Parmentier},
  \textsc{N.~Roch},  \textsc{J.~Viennot},  \textsc{G.~F{\`e}ve},
  \textsc{B.~Huard},  \textsc{C.~Mora},  \textsc{A.~Cottet},  and
  \textsc{T.~Kontos}\iffalse Coupling a quantum dot, fermionic leads and a
  microwave cavity on-chip\fi,
 \jr{Phys. Rev. Lett.} \textbf{107}, 256804 (2011).


\bibitem{Jaynes63:89}% article
 \textsc{E.\,T. Jaynes} and  \textsc{F.\,W. Cummings}\iffalse Comparison of
  quantum and semiclassical radiation theory with application to the beam
  maser\fi,
 \jr{Proc. IEEE.} \textbf{51}, 89 (1963).


\bibitem{Amitabh93:2276}% article
 \textsc{A.~Joshi} and  \textsc{S.\,V. Lawande}\iffalse Generalized
  jaynes-cummings models with a time-dependent atom-field coupling\fi,
 \jr{Phys. Rev. A} \textbf{48}, 2276 (1993).


\bibitem{Jonasson2011:01}% article
 \textsc{O.~Jonasson},  \textsc{C.\,S. Tang},  \textsc{H.\,S. Goan},
  \textsc{A.~Manolescu},  and  \textsc{V.~Gudmundsson}\iffalse Quantum
  magneto-electrodynamics of electrons embedded in a photon cavity\fi,
 \jr{New Journal of Physics} \textbf{14}, 013036 (2012).


\bibitem{Moldoveanu10:155442}% article
 \textsc{V.~Moldoveanu},  \textsc{A.~Manolescu},  \textsc{C.\,S. Tang},  and
  \textsc{V.~Gudmundsson}\iffalse Coulomb interaction and transient charging of
  excited states in open nanosystems\fi,
 \jr{Phys. Rev. B} \textbf{81}, 155442 (2010).


\bibitem{Gudmundsson10:205319}% article
 \textsc{V.~Gudmundsson},  \textsc{C.\,S. Tang},  \textsc{O.~Jonasson},
  \textsc{V.~Moldoveanu},  and  \textsc{A.~Manolescu}\iffalse Correlated
  time-dependent transport through a two-dimensional quantum structure\fi,
 \jr{Phys. Rev. B} \textbf{81}, 205319 (2010).


\bibitem{Gudmundsson12:1109.4728}% article
 \textsc{V.~Gudmundsson},  \textsc{O.~Jonasson},  \textsc{C.\,S. Tang},
  \textsc{H.\,S. Goan},  and  \textsc{A.~Manolescu}\iffalse Time-dependent
  transport of electrons through a photon cavity\fi,
 \jr{Phys. Rev. B} \textbf{85}, 075306 (2012).


\bibitem{Moldoveanu09:073019}% article
 \textsc{V.~Moldoveanu},  \textsc{A.~Manolescu},  and
  \textsc{V.~Gudmundsson}\iffalse Geometrical effects and signal delay in
  time-dependent transport at the nanoscale\fi,
 \jr{New Journal of Physics} \textbf{11}(7), 073019 (2009).


\bibitem{Gudmundsson09:113007}% article
 \textsc{V.~Gudmundsson},  \textsc{C.~Gainar},  \textsc{C.\,S. Tang},
  \textsc{V.~Moldoveanu},  and  \textsc{A.~Manolescu}\iffalse Time-dependent
  transport via the generalized master equation through a finite quantum wire
  with an embedded subsystem\fi,
 \jr{New Journal of Physics} \textbf{11}(11), 113007 (2009).


\bibitem{Nakajima58:948}% article
 \textsc{S.~Nakajima},
 \jr{Prog. Theor. Phys.} \textbf{20}, 948 (1958).


\bibitem{Zwanzig60:1338}% article
 \textsc{R.~Zwanzig},
 \jr{J. Chem. Phys.} \textbf{33}, 1338 (1960).


\end{thebibliography}
\providecommand{\WileyBibTextsc}{}
\let\textsc\WileyBibTextsc
\providecommand{\othercit}{}
\providecommand{\jr}[1]{#1}
\providecommand{\etal}{~et~al.}

\end{document}